\documentclass[runningheads]{llncs} 
\usepackage[utf8]{inputenc}
\usepackage[english]{babel}
\usepackage{graphicx}
\usepackage{xcolor}
\usepackage{amssymb}
\usepackage[normalem]{ulem}
\usepackage[color=yellow!20]{todonotes}
\usepackage{soul}
\usepackage{hyperref}
\usepackage{paralist}
\usepackage{mathrsfs}
\usepackage{listings}

\definecolor{lightgray}{rgb}{0.9,0.9,0.9}
\let\subparagraph\paragraph

 \usepackage{titlesec}
 \titleformat{\paragraph}[runin]
   {\normalfont\normalsize\bfseries}{\theparagraph.}{1em}{}[.]

\usepackage[paperheight=23.5cm,paperwidth=15.5cm,text={12.2cm,19.3cm},centering]{geometry} 

\usepackage[innerleftmargin=5pt,innerrightmargin=5pt]{mdframed}

\usepackage[lambda,advantage,asymptotics,adversary,complexity,sets,keys,ff,primitives,operators,probability,logic,mm,events]{cryptocode}

\usepackage{bm}

\usepackage[capitalise]{cleveref}
\usepackage{caption}
\usepackage{subcaption}
\usepackage[rightcaption]{sidecap}

\usepackage[mathscr]{euscript}

\usepackage[all]{nowidow}

\definecolor{darkred}{rgb}{0.7,0,0}
\definecolor{lightblue}{rgb}{0.85,0.85,1}
\definecolor{bananamania}{rgb}{0.98, 0.91, 0.71}

\newcommand{\commentfont}{\scriptsize}

\DeclareRobustCommand{\michals}[2]  {{\color{darkred}\sethlcolor{bananamania}\hl{\commentfont\textbf{Michal #1:}\ #2}}}

\newcommand{\comment}[1]{}

\let\vec\bm
\newcommand{\alg}[1]{\mathsf{#1}}
\newcommand{\contract}[1]{\mathtt{#1}}

\newcommand{\com}{\alg{Com}}
\newcommand{\LAN}{\mathbf{L}}
\newcommand{\REL}{\mathbf{R}}
\newcommand{\crs}{\vec{\mathsf{crs}}}
\newcommand{\td}{\vec{\mathsf{td}}}
\newcommand{\smallset}[1] {\{#1\}}
\newcommand{\inp}{\mathsf{x}}
\newcommand{\wit}{\mathsf{w}}

\newcommand{\game}[1]{\mathsf{#1}}
\mathchardef\mhyphen="2D 

\newcommand{\IKCCA}{\game{IK\mhyphen CCA}}
\newcommand{\INDCCA}{\game{IND\mhyphen CCA2}}
\newcommand{\party}[1]{\mathcal{#1}}
\DeclareMathAlphabet\mathbfcal{OMS}{cmsy}{b}{n}

\newcommand{\softwareComponent}[1]{\mathtt{#1}}

\newcommand{\zethgithub}{\url{https://github.com/clearmatics/zeth}}
\newcommand{\zeth}{\textup{\textnormal{\textsf{ZETH}}}}
\newcommand{\variable}[1]{\mathsf{#1}}
\newcommand{\alice}{\party{A}}
\newcommand{\bob}{\party{B}}
\newcommand{\charlie}{\party{C}}
\newcommand{\addr}{\variable{addr}}
\newcommand{\AddressRegistry}{\contract{Addr}}
\newcommand{\Mixer}{\contract{Mxr}}
\newcommand{\MerkleTree}{\contract{MTree}}
\newcommand{\Verifier}{\contract{Vrf}}
\newcommand{\currency}[1]{\textsf{#1}}
\newcommand{\zethNote}{\currency{zethNote}}
\newcommand{\zethNotes}{\currency{zethNote}s}
\newcommand{\zethAddress}{\textsf{zethAddress}}
\newcommand{\zethAddresses}{\textsf{zethAddresses}}
\newcommand{\znote}{\mathbf{z}}
\newcommand{\snark}{\alg{\Psi}}
\newcommand{\tx}{\variable{tx}}
\newcommand{\txmix}{\tx_{\mix}}

\newcommand{\mix}{\textnormal{\ensuremath{\alg{Mix}}}}

\newcommand{\receive}{\alg{Receive}}
\newcommand{\ledger}{\mathscr{L}}

\newcommand{\cxx}{\mathsf{c}}
\newcommand{\balance}{\game{BAL}}

\newcommand{\val}{v}

\newcommand{\ENC}{\alg{E}}
\newcommand{\oracleo}{\alg{O}}

\newcommand{\cm}{\variable{cm}}
\newcommand{\sn}{\variable{sn}}
\newcommand{\rt}{\variable{rt}}
\newcommand{\new}{\variable{new}}
\newcommand{\old}{\variable{old}}

\newcommand{\setsender}{\mathbfcal{S}}
\newcommand{\recipient}{\party{R}}
\newcommand{\setrecipient}{\mathbfcal{R}}

\newcommand{\mixindcommand}[1]{\mathbf{#1}}

\newcommand{\mixAdv}{\mixindcommand{Mix}}

\newcommand{\insertAdv}{\mixindcommand{Insert}}
\newcommand{\createAddressAdv}{\mixindcommand{CreateAddress}}
\newcommand{\ADDR}{\variable{ADDR}}

\newcommand{\coinsIn}{N}
\newcommand{\coinsOut}{M}
\newcommand{\NOTE}{\variable{NOTE}}

\newcommand{\commenti}[1]{{\footnotesize /\!\!/ #1}}
\newcommand{\smallvec}[1]{\smallset{#1}}
\newcommand{\sha}{\mathtt{sha256}}

\newcommand{\range}[2]{[#1 .. #2]}
\newcommand{\receiveAdv}{\mixindcommand{Receive}}
\newcommand{\TRNM}{\game{TR\mhyphen NM}}
\newcommand{\oledger}{\oracleo_{\ledger}}
\newcommand{\vin}{\mathsf{in}}
\newcommand{\vout}{\mathsf{out}}
\newcommand{\project}[1]{\texttt{#1}}

\newcommand{\COMM}{\alg{\Gamma}}

\newcommand{\PairBaseGas}{\variable{PAIRING\_BASE\_GAS}}
\newcommand{\PairPerPointGas}{\variable{PAIRING\_PER\_POINT\_GAS}}

\newcommand{\ECADDGas}{\variable{ECADD\_GAS}}
\newcommand{\ECMULGas}{\variable{ECMUL\_GAS}}

\newcommand{\EthTxObjSender}{\variable{EthTxObj.sender}}
\newcommand{\EthTxObjValue}{\variable{EthTxObj.value}}
\newcommand{\SerialNumberList}{\variable{SerialNumberList}}
\newcommand{\MerkleRootList}{\variable{MerkleRootList}}
\newcommand{\MerkleTreeVar}{\variable{MerkleTree}}
\newcommand{\insertFunc}{\alg{Insert}}
\newcommand{\insertLeaf}{\alg{InsertLeaf}}
\newcommand{\contain}{\alg{Contain}}
\newcommand{\isEqual}{\alg{IsEqual}}
\newcommand{\sendValue}{\alg{SendValue}}
\newcommand{\getMerkleRoot}{\alg{GetMerkleRoot}}
\newcommand{\broadcastCiphertext}{\alg{BroadcastCiphertext}}
\newcommand{\broadcastMerkleRoot}{\alg{BroadcastMerkleRoot}}

\newcommand{\proofGenerator}{\softwareComponent{PG}}
\newcommand{\proofConsumer}{\softwareComponent{PC}}
\newcommand{\getVerificationKey}{\alg{GetVerificationKey}}

\newcommand{\getProof}{\alg{GetProof}}

\newcommand{\statesc}[1]{\mathscr{#1}}
\newcommand{\stateMixer}{\statesc{M}}

\newcommand{\setznotes}{\mathbf{Z}}

\newcommand{\master}{\variable{mstr}}
\newcommand{\specialized}{\variable{spec}}
\newcommand{\crsmaster}{\crs^{\master}}
\newcommand{\crsspec}{\crs^{\specialized}}
\newcommand{\ffam}{\mathscr{H}}

\newcommand{\funspace}[1]{\mathscr{#1}}
\newcommand{\keyspace}{\funspace{K}}
\newcommand{\aspace}{\funspace{A}}
\newcommand{\bspace}{\funspace{B}}

\newcommand{\mixinput}{
	\pi,\allowbreak
	\rt,\allowbreak
	v_{\vin},\allowbreak
	v_{\vout},\allowbreak
	\smallvec{\cm^{\new}_{i}}_{i = 1}^{\coinsOut},\allowbreak
	\smallvec{\sn_{i}}_{i = 1}^{\coinsIn},\allowbreak
	\smallvec{\cxx_i}_{i = 1}^{\coinsOut}
}
\newcommand{\verifyinput}{
	\pi,\allowbreak
	\rt,\allowbreak
	v_{\vin},\allowbreak
	v_{\vout},\allowbreak
	\smallvec{\cm^{\new}_{i}}_{i = 1}^{\coinsOut},\allowbreak
    \smallvec{\sn_{i}}_{i = 1}^{\coinsIn}
}
\newcommand{\WEI}{\texttt{Wei}}
\newcommand{\ETH}{\texttt{ETH}}
\newcommand{\ETHone}{\texttt{ETH\,1.0}}

\newcommand{\babyzoe}{\project{babyZoE}}
\newcommand{\validProof}{\variable{valid\_proof}}
\newcommand{\CRSGenerator}{\softwareComponent{CG}}
\renewcommand{\AA}{\mathbf{A}}
\newcommand{\BB}{\mathbf{B}}
\renewcommand{\CC}{\mathbf{C}}
\definecolor{ballblue}{rgb}{0.13, 0.67, 0.8}

\newcommand{\miximus}{\project{Miximus}}
\newcommand{\mobius}{\project{M\"obius}}
\newcommand{\quisquis}{\project{Quisquis}}
\newcommand{\aztec}{\project{AZTEC}}
\newcommand{\zether}{\project{Zether}}
\newcommand{\RELGEN}{\mathcal{R}}
\newcommand{\mkPath}{\variable{mkPath}}
\newcommand{\cmAddr}{\variable{cmAddr}}
\newcommand{\enforce}{\variable{e}}
\newcommand{\RELCIRC}{\REL^{\project{z}}}

\makeatletter
\renewcommand\subsubsection{\@startsection{subsubsection}{3}{\z@}%
	{-18\p@ \@plus -4\p@ \@minus -4\p@}%
	{0.5em \@plus 0.22em \@minus 0.1em}%
	{\normalfont\normalsize\bfseries}}
\makeatother
\title{\zeth{}: On Integrating Zerocash on Ethereum}
\author{Antoine Rondelet \and Michal Zajac}
\institute{
  Clearmatics, UK \newline
  \email{\{ar$\|$michal.zajac\}@clearmatics.com}}
\date{\today}

\begin{document}
\maketitle

\begin{abstract}

    Transaction privacy is a hard problem on an account-based blockchain such as Ethereum. While Ben-Sasson et al. presented the Zerocash protocol~\cite{zerocash-paper} as a decentralized anonymous payment (DAP) scheme standing on top of Bitcoin, no study about the integration of such DAP on top of a ledger defined in the account model was provided. In this paper we aim to fill this gap and propose \zeth{}, an adaptation of Zerocash that can be deployed on top of Ethereum without making any change to the base layer. Our study shows that not only \zeth{} could be used to transfer Ether, the base currency of Ethereum, but it could also be used to transfer other types of smart contract-based digital assets.
We propose an analysis of \zeth{}'s privacy promises and argue that information leakages intrinsic to the use of this protocol are controlled and well-defined, which makes it a viable solution to support private transactions in the context of public and permissioned chains.

\end{abstract}

\section{Introduction}
The two cryptocurrencies with the biggest market capitalization are Bitcoin~\cite{bitcoin-whitepaper} and Ethereum~\cite{ethereum-whitepaper}. One of the crucial differences between them is the way they store information. While Bitcoin relies on the \emph{UTXO model}, which stores the list of all unspent transaction outputs, Ethereum is \emph{account based}, that is, it keeps track of accounts' balances.
Although neither currencies provide strong privacy properties to their users, Zerocash~\cite{zerocash-paper} practically solved this issue on Bitcoin-like systems.
Unfortunately, integrating Zerocash on Ethereum is not described in the paper and does not seem to be feasible out of the box.

In this paper we present \zeth{} -- a variation of Zerocash that allows for secure and private payments on Ethereum. We claim that \zeth{} can easily be used with any ERC20, or ERC223 token\footnote{For discussion about these token standards check \url{https://github.com/ethereum/eips/issues/20} and \url{https://github.com/ethereum/eips/issues/223}.}, and can be used along any Ethereum client.
Moreover, \zeth{} uses minimal trust assumptions and no asymmetry is introduced in the  system, namely, no trusted or special node or role  is required to satisfy our privacy promises.
Finally, we provide a proof of concept implementation of \zeth{}\footnote{\zethgithub}.

\subsection{Privacy limitations of Ethereum}
The Ethereum protocol provides an extension to Bitcoin by enabling execution of arbitrary computation on-chain on the so-called Ethereum Virtual Machine (EVM), which is Turing complete.
Ethereum can be seen as a~distributed state machine, where every transaction triggers a state transition. These transitions modify the underlying state that keeps track of the account balances and contracts' storage. As a consequence, every transaction done from an account is publicly recorded on the blockchain and modifies the public state.

Using smart contracts as a way to provide privacy preserving state transitions appears to be an appealing idea.
Unfortunately, such solutions are not perfect and come with several drawback due to the gas model that was introduced to bound the computation of each state transition and keep the system running.

In the gas model each operation executed on the EVM uses a number of gas units and the minimal gas cost of a transaction is $21\,000$ gas.
An Ethereum user who wants to run a smart contract function has to specify the amount of gas units to be used for the function execution, along with the price in \WEI{}\footnote{\WEI{} is the base unit (or smallest denomination) of Ether.} they are willing to pay per gas unit (\emph{gas price}).
These two values are used to determine the maximum cost of the transaction the sender is willing to pay. Denote this upper bound by $u$\,\WEI.
When the transaction is submitted to the system and gets processed, $u$\,\WEI{} are secured at the user’s account.
The effective gas usage of the transaction, denoted $e$\,\WEI{}, is computed as the transaction is executed. The unused gas, namely $u - e$\,\WEI{}, is returned to the sender.
If the execution of a smart contract function runs out of gas\footnote{The amount of gas required to execute the function is higher than the amount of gas paid by the user.}, the modifications to the state carried out during the execution of this function are reverted, the miner\footnote{Node that reaches consensus over the Ethereum state and that produces blocks by executing a set of transactions submitted by network users.} collects a fee corresponding to the gas used and the transaction sender does not get any \WEI{} back.

Crucially, enforcing the payment of gas for every transaction processed on Ethereum prevents malicious users from saturating the network with dummy transactions and restrain adversaries from deploying smart contracts representing never-ending state transitions, e.g.~infinite loops.

Since paying for transaction fees directly alters Ethereum accounts, and thus modifies the public state, everybody can tell the nature of a transaction, what data it carries and how much gas was paid by the transaction sender.
As a consequence, having a smart contract-based private payment solution still requires users to interact with the contract and pay for its execution. This translates in information leakage and can disclose a potential relation between the users interacting with the contract.

Furthermore, implementing a smart contract-based privacy payment solution can be quite expensive. In fact, since users need to pay for the execution of the contract, any expensive on-chain computation leads to high amount of gas to be paid by the smart contracts callers. The expensiveness of privacy on Ethereum has already been pointed out by the community \cite{privacy-expensive}, and should hopefully improve in the following releases of the platform.

Finally, the gas model inherent to Ethereum implies that several privacy enhancing techniques, initially described in the context of Bitcoin, cannot be applied to Ethereum (at least on a Byzantium fork). One of such concept is the notion of stealth addresses~\cite{stealthaddress}.
Using the stealth addresses mechanism to generate new identities on the network cannot be used on Ethereum because of the need to pay gas for every transaction. That is, if a user tries to send a transaction from a stealth address, the transaction will be considered invalid, because the balance associated with the address is $0$. Consequently, no transactions can be initiated from it.
A naive solution to this issue would be to fund the stealth address from another, potentially publicly known, address. Unfortunately, such action may introduce a link between the funded address and the newly generated stealth address\footnote{Depending on the system, it might be feasible to rely on a trusted third party to fund the newly generated addresses. Nevertheless, this comes at the cost of more centralization and requires some degree of coordination which can affect the off-chain communication complexity.}.

\subsection{\zeth{} protocol}\label{sec:zeth-protocol}
\zeth{} implements a private payment mechanism by running a mixing smart contract on which members of the Ethereum network willing to transact privately deposit their funds in exchange for \zethNotes{}. The notes are stored on the mixer as commitments and can be privately transferred among users of the network.
Importantly, once the funds enter the mixer it is not necessary to withdraw them since \zethNotes{} can be used for future payments within the smart contract.
This property essentially increases the security and privacy of the proposed solution. Intuitively, the longer the funds are circulating in the mixer, the harder it is to connect them with their previous owners.

In order to provide the same security guarantees as UTXO systems operating on public data and keep the system sound (funds are bound to an owner and cannot be double spent), \zeth{} follows the construction detailed in Zerocash.
	First of all, the notes themselves are not transferred between the parties. Rather, destruction of some $\coinsIn$ notes belonging to the sender allows for creation of other $\coinsOut$ notes possessed and controlled by the recipient.

	Secondly, with an overwhelming probability, each note has a distinct and pseudorandom \emph{serial number} $\sn$ (also called \emph{nullifier}). When the note is spent (destroyed), its fixed serial number is revealed on the network to ensure that this note will not be spent a second time. Moreover, the pseudorandomness of this number assures that it is infeasible to connect previously published commitments with the spent notes.

	Thirdly, the creation of new notes comes with the publication of a list of commitments that hide all new notes' attributes, e.g.~the value, public key of the recipient, and nonces $\varrho$, $r$. Encryption of these attributes is broadcast. The ciphertexts are encrypted using the recipients' encryption keys, such that they can easily learn hidden attributes necessary to later spend the notes.

	Last but not least, correctness of these operations is provided by a succinct zero-knowledge  proof of knowledge (cf.~\cref{sec:overview-snarks}), which assures that
	the sender is the rightful owner of the notes,
	the commitments to the spent notes were previously published, and
	the transaction is balanced, i.e.~the value of notes input to the transaction equals the value output. We note here that some inputs and outputs may be public.
	An overview of the protocol is provided in \cref{sec:overview-protocol} and more formal definitions of its security is shown in \cref{sec:security-guarantees}.
  Importantly, since the proof verification procedure is deterministic it can be executed on-chain.

\begin{figure}
	\centering
	\includegraphics[width=1\textwidth]{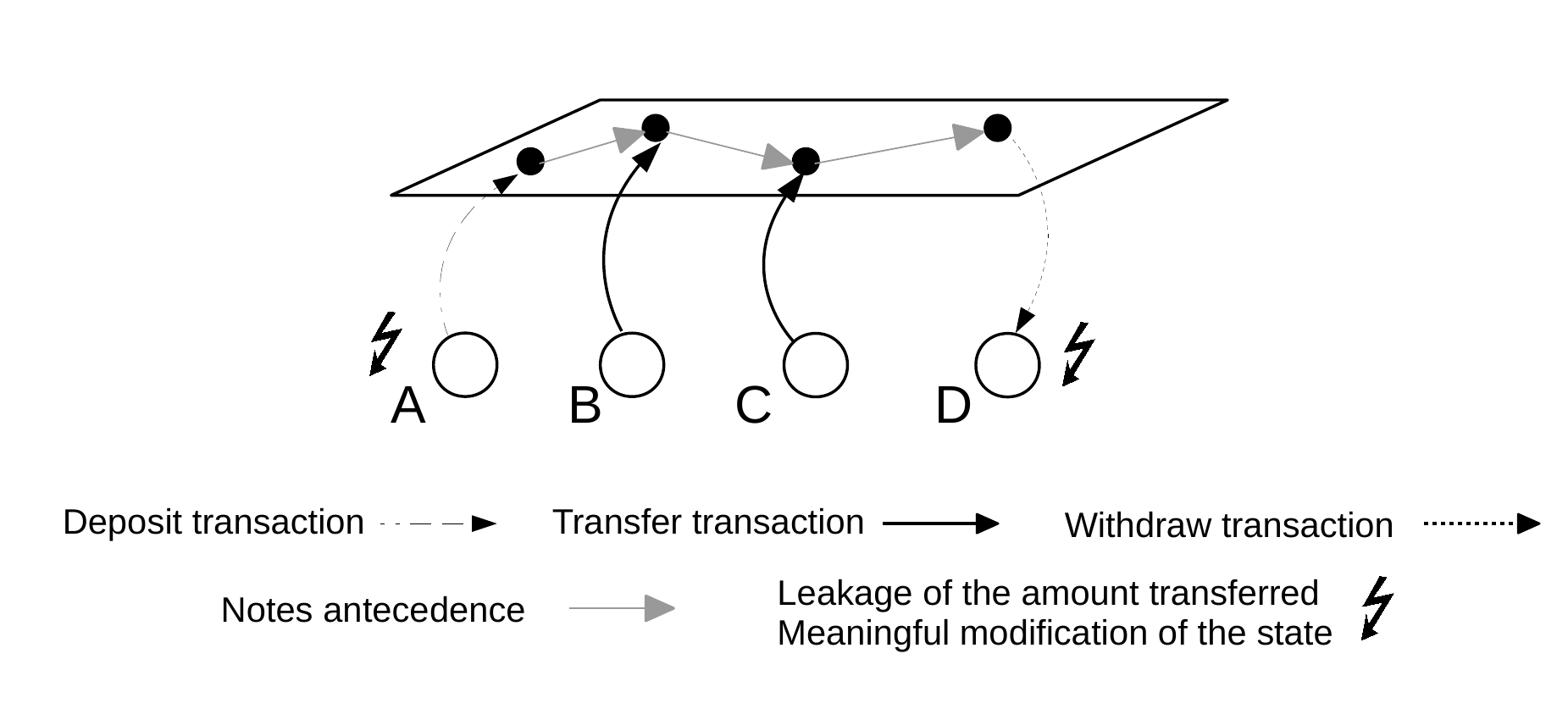}
	\caption{\zeth{} abstraction.}
	\label{fig:zeth-abstraction}
\end{figure}

\subsection{Security of \zeth{}}
Since each call to the mixer contract leaks the caller who pays the gas for the smart contract function execution, the list of users of the system is public.
That is, each transfer is accompanied by a modification of the Ethereum state. Especially, the balance of the caller in decremented by the gas cost of the transaction.
On the other hand, the recipient is fully hidden even when they want to use the funds by passing them on (i.e.~when the recipient effectively becomes the sender) or withdrawing them from the mixer, cf.~\cref{fig:zeth-abstraction}. While calling the mixer contract reveals that a party has funds to transfer and can expose the fact that they were recipient of a payment in the past; it is not possible to determine the payment they received.
Thus, despite the lack of obfuscation of users, we claim that \zeth{} is an appropriate solution to blur the transaction graph, and thus, hide relationships between senders and recipients, transactions' value, along with the wealth of each user of the mixing service.

To make a payment, the sender calls the contract by executing the $\mix$ function of the mixer smart contract and publishes the serial numbers of the input \zethNotes{}, the ciphertexts of the new notes, and the transaction's zero-knowledge proof of knowledge.
The encrypted \zethNotes{} are integrated as a part of the logs of the transaction and the state of the smart contract is changed since new commitments are appended and serial numbers are revealed.

We leverage the ability to emit events in a smart contract to implement an encrypted broadcast used to make sure that the sender and the recipient never need to communicate.
In addition to that, we build on the fact that state transitions corresponding to mixer contract calls do not reveal the same amount of information as state transitions triggered by plain Ethereum transactions.
In fact, the state transition encoded by the mixer's function results in the modification of the smart contract state (commitments and serial numbers are appended) along with some gas being paid by the smart contract caller.
Nonetheless, plain Ethereum transactions (e.g.~Alice pays 2\,\ETH{} to Bob) modify the public state which leaks the relationship of the sender and recipient along with the value of the payment.

\begin{figure}
    \centering
    \begin{subfigure}{0.4\textwidth}
    	\includegraphics[width=1\textwidth]{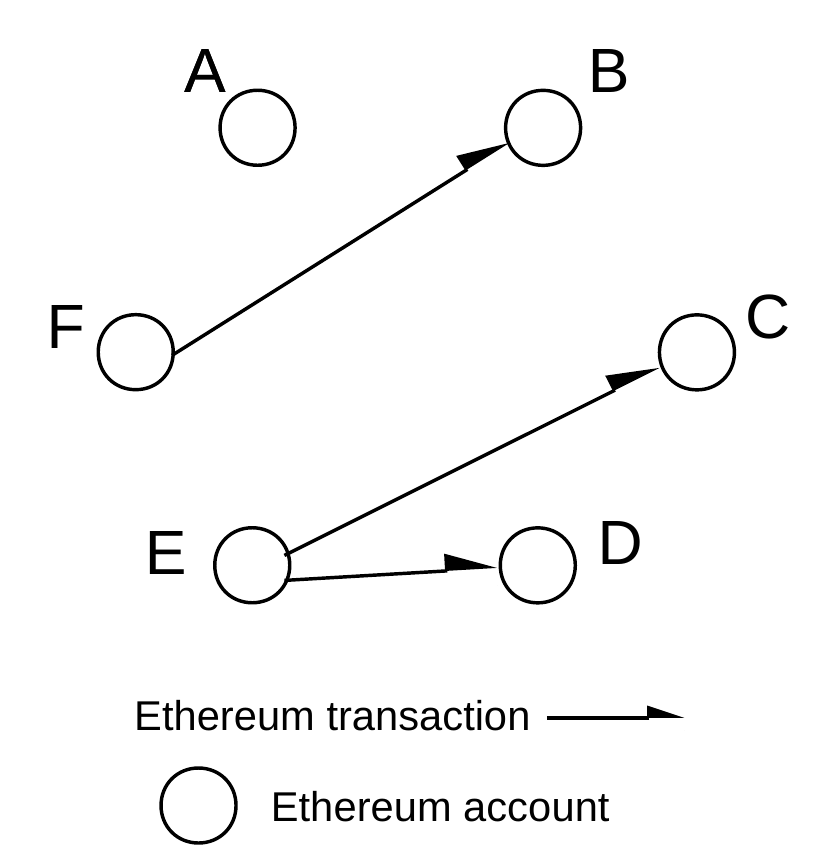}
    	\caption{Transaction graph without \zeth{}.}
    	\label{fig:transaction-graph-no-mixer}
    \end{subfigure}
	\hspace*{0.7cm}
    \begin{subfigure}{0.4\textwidth}
   		\includegraphics[width=1\textwidth]{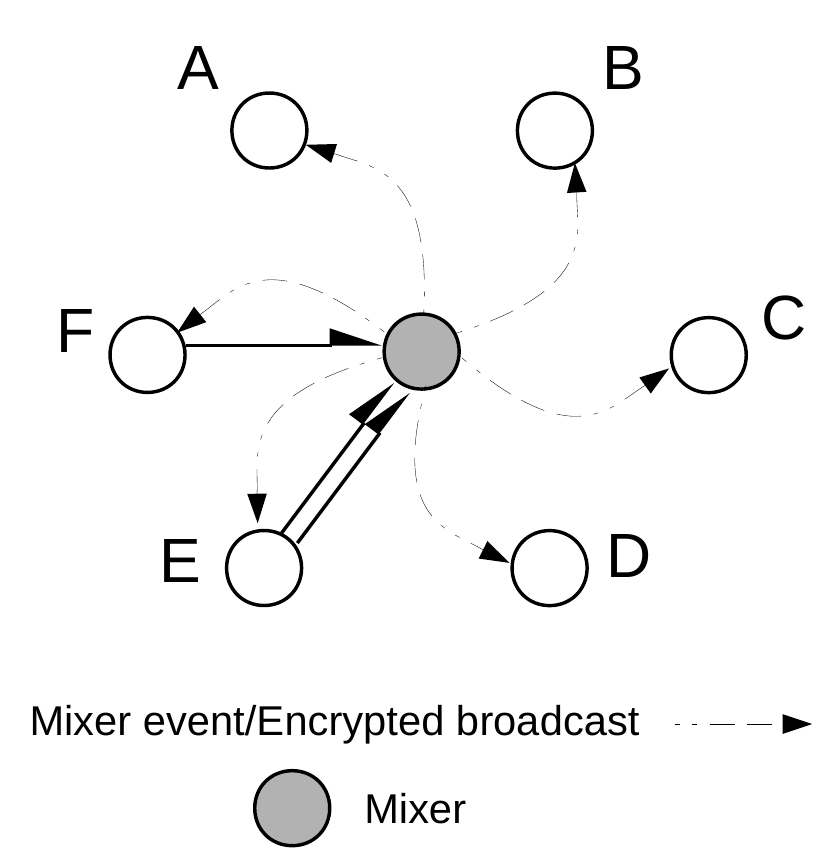}
   		\caption{Transaction graph using \zeth{}.}
   		\label{fig:transaction-graph-yes-mixer}
    \end{subfigure}
	\caption{Transaction graph without and with \zeth{}. Using the mixer changes the topology of the transaction graph replacing user-to-user edges with user-to-mixer and mixer-to-all-users ones.}
	\label{fig:transaction-graph}
\end{figure}

In that sense, the use of \zeth{} changes the topology of the transaction graph.
More precisely, let the graph's nodes represent Ethereum accounts and its (directed) edges correspond to transactions and messages between them. As shown in \cref{fig:transaction-graph-no-mixer}, each pair of transacting nodes are easy to identify.
However, when transactions go through \zeth{}, the mixer smart contract serves as an intermediary. Thus, instead of account-to-account edges, the graph is now made of account-to-mixer or mixer-to-all-accounts (encrypted broadcast) edges that obfuscate the relationship between users of the system, see \cref{fig:transaction-graph-yes-mixer}.

From the security point of view, it is important to notice that \zeth{}, using techniques introduced by Zerocash, allows for an arbitrary denomination of notes.
Although some solutions aim to provide private transactions of standard/predefined denominations, adding such a constraint has several drawbacks that can take users away from the recommended behaviors (see \cref{sec:best-practices}), or even render the entire system unusable. Furthermore, previous work~\cite{mobius-analysis} have shown that smart contract-based solutions using predefined denominations could eventually lead to several exploitable data leakages that could, in fine, undermine the privacy of its users.
\comment{
  Despite Ethereum's ability to perform payments of arbitrary denominations, it is possible to add more constraints on the system and design a solution for private transactions of standard/predefined denominations.
  Nonetheless, adding such a constraint has several drawbacks that can take users away from the recommended behaviors (see \cref{sec:best-practices}), or even render the entire system unusable. Furthermore, some pieces of work~\cite{mobius,mobius-analysis} have shown that smart contract-based solutions using predefined denominations could eventually lead to several exploitable data leakages that could, in fine, undermine the privacy of its users.
  Zerocash~\cite{zerocash-paper} tackles the issue of predefined denomination coins. It leverages zkSNARKs in order to allow to \emph{pour} the value of a set of coins into another set of newly generated coins. This ability is key and allows our mixing contract to support payments of arbitrary denominations just like the native platform.
}

\subsection{Divergence from Zerocash}
As mentioned, Zerocash has initially been designed to work as a fork of the Bitcoin blockchain. Since Ethereum and Bitcoin differ on many aspects, we needed to modify the Zerocash protocol in order to reflect the diversions of these two blockchains.

While most of our construction is based on the protocol described in the Zerocash paper \cite{zerocash-paper}, our protocol presents weaker privacy guarantees than the original design. In fact, users of the \zeth{} mixing service are not obfuscated.

Furthermore, we decided to leverage smart contract events as a way to implement an encrypted broadcast mechanism to exchange the \zethNotes{} between users of the system. This ability, natively supported by Ethereum, is used to provide an additional layer of privacy for the system users.

Finally, we decided to implement a single function on the mixer smart contract, that supports both \emph{minting} and \emph{pouring} coins. This comes with the drawback of requiring a zero-knowledge proof when one makes a deposit into the mixer (and \emph{mint} \zethNotes{}), but also comes with the benefit of a more flexible mixer and a better obfuscation of funds.

\subsection{Related work}\label{sec:previous-work}

Before we present our contribution to privacy on Ethereum, it is worth mentioning related works presented in the past.
A first step integration of Zerocash and Ethereum was made in 2016 as a joint effort between members of Zcash, Ethereum Foundation and Cornell University. The resulting code of this project called \babyzoe{} is available on Github\footnote{\url{https://github.com/zcash-hackworks/babyzoe}}.

Later, barryWhiteHat proposed a project \miximus{}\footnote{\url{https://github.com/barryWhiteHat/miximus}} that was used as a starting point for the first version of our proof of concept before adapting Zcash \project{Sprout} for the second version of the PoC.
Finally, a draft study of the integration of Zerocash on Ethereum has already been written and published on Github\footnote{\url{https://github.com/AntoineRondelet/zerocash-ethereum}} a few months ago. This paper formalizes, improves and presents an implementation of this initial work.

Recently, Meiklejohn et al.~\cite{mobius} proposed a ring signature-based tumbler called \mobius{}\footnote{\url{https://github.com/clearmatics/mobius}}.
This solution, aiming to enhance privacy on Ethereum, leverages \emph{linkable ring signatures} coupled with stealth addresses and stealth keys to enable private transactions while preventing double spending.
While the solution seems to be quite practical in regard of gas consumption and off-chain communication complexity, it only supports transactions of standard denominations and present some latency issues. Moreover, since the use of stealth addresses is not compatible with the current version of Ethereum, deployment of this solution on existing systems raises some challenges. However, incoming updates of the Ethereum network could pave the way for an easier use of this approach.

Likewise, a new design for anonymous cryptocurrencies called \quisquis{}~\cite{DBLP:journals/iacr/FauziMMO18} has been proposed. This solution relies on the notion of \emph{updatable public keys}, and keeps track of users' wealth using an account structure. The balance of a given account is kept in an obfuscated, committed form and can be updated using the homomorphic property of the commitment scheme.
Being account based, \quisquis{} presents some advantages regarding scalability. Nevertheless, the potentially (very) long list of a user's updated keys may present a challenge in term of key management and several issues regarding transactions asynchrony are still to be solved.

More recently, the \aztec{} protocol was published~\cite{AZTEC}. This efficient privacy enhancing protocol seating on top of Ethereum uses a form of \emph{special purpose} zero-knowledge proofs leveraging Boneh-Boyen signatures to create a commitment scheme equipped with efficient range proofs. While \aztec{} does not rely on any zkSNARK, it also comes with the drawback of requiring a trusted setup, which is expected to be carried out via a multiparty computation ceremony.

Last but not least, B\"unz et al.~proposed \zether{}~\cite{zether}, a private payment mechanism atop of Ethereum. Similarly to \quisquis{}, \zether{} is account-based, what could potentially ease the interaction with other Ethereum-based smart contracts. In support of this claim, the authors show several applications of their mechanism like sealed-bid auctions, payment channels, stake voting and privacy-preserving proofs of stake.
Nevertheless, the design of \zether{} comes with some drawbacks, one of them being the cost of a transaction. Furthermore, the size of a zero-knowledge proof $\pi$ in a \zether{} transfer depends on the size of the anonymity set, which is preferably as big as possible. Also, since the statement for $\pi$ depends on the sender's account balance, the account has to be locked to make sure the proof is acceptable.

\section{Preliminaries}
\subsection{Notations}
Let PPT denote probabilistic polynomial time and let NUPPT denote non-uniform PPT. Let $\secpar \in \NN$ be the security parameter, in practice we set e.g.~$\secpar = 128$. All adversaries are stateful.
An adversary and all parties they control are denoted by letter $\adv$.
We write $\negl[]$ or $\poly[]$ to denote a negligible or polynomial function resp. We say that the adversary wins game $\game{GAME}$ if they make $\game{GAME}$ return $1$.
We say that $\adv$ wins $\game{GAME}$ with negligible advantage if $\advantage{\game{GAME}}{\adv, \game{GAME}} = 2 \abs{ \prob{\game{GAME} = 1} - 1/2} \leq \negl$.
In this paper we follow the conventions of \cite{zerocash-paper} and use $\sha$ as a hash function inside the Merkle tree, as a commitment scheme, and as a pseudorandom function.

\subsection{Cryptographic primitives}
\paragraph{Collision resistant hash function family}

Informally speaking, a collision resistant hash function family $\ffam$ is a collection of functions such that no PPT algorithm given random $h \in \ffam$ can find $x$ and $x'$, such that $h (x) = h (x')$ and $x \neq x'$ with non-negligible probability.
In the real world it is usually assumed that the hash function family consists of a single function, e.g.~$\sha$. For a formal definition see, e.g.,~\cite{DBLP:books/cu/Goldreich2001}.

\paragraph{Key-private and CCA2-secure encryption scheme}
An encryption scheme $\ENC$ consists of the following three algorithms:
\begin{description}
  \item[$\kgen(\secparam)$:] given a security parameter as input returns a public key $\pk$ and a secret key $\sk$;
  \item[$\enc_{\pk}(m, r)$:] on input $m$ outputs $c$, a ciphertext that encrypts plaintext $m$ under public key $\pk$ and randomness $r$;
  \item[$\dec_{\sk}(c)$:] given ciphertext $c$ and secret key $\sk$ returns plaintext $m$ such that for any $r$ holds $\dec_{\sk} (\enc_{pk} (m, r)) = m$.
\end{description}

\begin{figure}
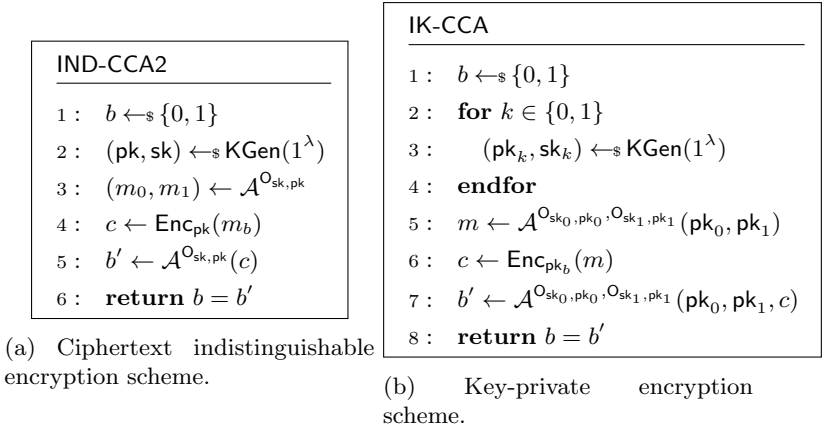

  \centering
	\begin{subfigure}{0.4\textwidth}
		\centering
		\fbox{
			\procedure[linenumbering]{$\INDCCA$}{
				b \sample \bin\\
				(\pk, \sk) \sample \kgen(\secparam) \\
				(m_0, m_1) \gets \adv^{\oracleo_{\sk, \pk}}\\
				c \gets \enc_{\pk}(m_b)\\
				b' \gets \adv^{\oracleo_{\sk, \pk}} (c)\\
				\pcreturn b = b'
			}
		}
		\caption{Ciphertext indistinguishable encryption scheme.}
		\label{fig:indcca}
	\end{subfigure}
 	\begin{subfigure}{0.4\textwidth}
 	\centering
	\fbox{
		\procedure[linenumbering]{$\IKCCA$}{
			b \sample \bin\\
			\pcfor k \in \bin \\
			\pcind (\pk_k,\sk_k) \sample \kgen(\secparam)\\
			\pcendfor \\
			m \gets \adv^{\oracleo_{\sk_0, \pk_0}, \oracleo_{\sk_1, \pk_1}} (\pk_0, \pk_1)\\
			c \gets \enc_{\pk_b}(m)\\
			b' \gets \adv^{\oracleo_{\sk_0, \pk_0}, \oracleo_{\sk_1, \pk_1}} (\pk_0, \pk_1, c)\\
			\pcreturn b = b'
		}
	}
	\caption{Key-private encryption scheme.}
	\label{fig:ikcca}
\end{subfigure}
	\caption{We require that the encryption scheme we use is both $\INDCCA$ and $\IKCCA$-secure. That is, for any NUPPT adversary $\adv$, the probability that $\adv$ has non-negligible advantage in one of the games is negligible.}
  \label{fig:ikindcca}
\end{figure}

As usually, we require that the encryption scheme is ciphertext-indistinguishable ($\INDCCA$). That is, for any NUPPT adversary $\adv$ and sufficiently large $\secpar$
\[
\advantage{\INDCCA}{\adv, \INDCCA} \leq \negl \enspace,
\]
for the $\INDCCA$ game defined at \cref{fig:indcca}.
In this game the adversary $\adv$ is equipped with oracle access to an oracle $\oracleo$ that encrypts messages and decrypts ciphertexts submitted by $\adv$. Finally, the adversary provides two messages $m_0$ and $m_1$ and in response gets one of them (picked at random) encrypted in a ciphertext $c$. $\adv$ wins the game if they guess whether $c$ hides $m_0$ or $m_1$. Importantly, $\adv$ can query $\oracleo$ even after they learned $c$. However, the oracle answers $\bot$ if queried on the challenge ciphertext $c$.

In addition to be $\INDCCA$-secure, we also require $\ENC$ to be key-private.
Roughly speaking, $\ENC$ is called key-private~\cite{DBLP:conf/asiacrypt/BellareBDP01} if it hides the public key that was used to generate the ciphertext. Formally we require that for any NUPPT adversary $\adv$ and sufficiently large $\secpar$
\[
	\advantage{\IKCCA}{\adv, \IKCCA} \leq \negl \enspace,
\]
for the $\IKCCA$ game defined at \cref{fig:ikcca}.
In this game the adversary $\adv$ gets access to two encryption-decryption oracles each of them parameterized with a different pair of keys, either $(\pk_0, \sk_0)$ or $(\pk_1, \sk_1)$. Similarly to the $\INDCCA$ game, the adversary provides message $m$ and in response gets its encryption $c$ under key $\pk_0$ or $\pk_1$. $\adv$ wins the game if they guess whether $c$ was encrypted using $\pk_0$ or $\pk_1$. Importantly, $\adv$ can query their oracles even after they learned $c$. However, the oracles answer $\bot$ if queried on the challenge ciphertext $c$.

\paragraph{Pseudorandom function family}
Informally speaking, a pseudorandom function family
$\smallset{\prf_k: \aspace \to \bspace}_{k \in \keyspace}$
is a collection of functions such that for randomly chosen $k \in \keyspace$, function $\prf_{k}$ is indistinguishable from a random function that maps $\aspace$ to $\bspace$.
For a formal definition see, e.g.,~\cite{DBLP:books/cu/Goldreich2001}.

\paragraph{Commitment schemes}
A commitment scheme $\COMM$ is the compound of two algorithms $(\com, \verify)$ such that:
\begin{description}
  \item[$\com_{r}(m)$:] given message $m$ and randomness $r$, returns commitment $c$;
  \item[$\verify(c, m, r)$:] given commitment $c$, message $m$ and randomness $r$, checks whether $c = \com_{r} (m)$ and accepts iff that is the case.
\end{description}

We say that $\COMM$ is statistically binding if no adversary $\adv$, even computationally unbounded, can produce commitment $c$ and two openings $(m, r)$, $(m', r')$, such that $c = \com_{r}(m) = \com_{r'}(m')$.
Also, $\COMM$ is computationally hiding if for every $m, m'$, the probability distributions of $\com_{r} (m)$ and $\com_{r'}(m')$ are computationally indistinguishable over the choice of randomness $r, r'$.

\subsection{Brief overview of zkSNARKs}\label{sec:overview-snarks}
Let $\RELGEN$ be a relation generator, such that $\RELGEN (\secparam)$ returns a polynomial-time decidable binary relation $\REL = \smallset{(\inp, \wit)}$. Here, $\inp$ is the instance and $\wit$ is the witness.
We assume that $\secpar$ is explicitly deductible from the description of $\REL$.
Let $\LAN$ be an $\npol$ language corresponding to the relation $\REL$.
Roughly speaking, $\snark$ is a publicly verifiable
zero-knowledge Succint Non-interactive Argument of Knowledge (zkSNARK) for $\LAN$ if $\snark$ comports four PPT algorithms
$\kgen, \prover, \verifier, \simulator$ such that:
\begin{description}
    \item[CRS generator:] $\kgen$ is a PPT algorithm that takes security parameter $\secpar$ (presented in unary) as input, runs a one time setup routine, and outputs a \emph{common reference string} (CRS) $\crs$ along with a \emph{trapdoor} $\td$. The part of $\crs$ used for proving statements is sometimes called \emph{proving key} and denoted $\crs_\prover$, and the part necessary for their verification is called \emph{verification key} and denoted $\crs_\verifier$.
    \item[Prover:]  $\prover$ is a PPT algorithm that given $(\crs, \inp, \wit)$, such that $(\inp, \wit) \in \REL$, outputs an argument $\pi$.
    \item[Verifier:] $\verifier$ is a PPT algorithm that on input $(\crs, \inp, \pi)$ returns either $0$ (reject) or $1$ (accept).
    \item[Simulator:] $\simulator$ is a PPT algorithm that on input $(\crs, \td, \inp)$ outputs an argument $\pi$.
\end{description}

We require a proof system $\snark$ to have the following four properties (see, e.g.,~\cite{DBLP:journals/iacr/GrothM17} for their formal definition):
\begin{description}
  \item[Completeness:] $\snark$ is complete if an honest verifier accepts a proof made by an honest prover. That is, the verifier accepts a proof made for $(\inp, \wit) \in \REL$.
  \item[Knowledge soundness:] $\snark$ is knowledge-sound if from an acceptable proof $\pi$ for instance $\inp$ it is feasible for a specialized algorithm called extractor to extract a witness $\wit$ such that $(\inp, \wit) \in \REL$. (Note, knowledge soundness implies soundness.)
  \item[Zero knowledge:] $\snark$ is zero-knowledge if for any $\inp \in \LAN$ no adversary can distinguish a proof made by an honest prover equipped with $(\crs, \inp, \wit)$ from a proof made by the simulator equipped with $(\crs, \td, \inp)$ but no witness $\wit$.
  \item[Succinctness:] $\snark$ is succinct if the proof $\pi$ is sub-linear to the size of the instance and witness\footnote{By definition the proof can be of polynomial size to the length of the security parameter, however modern zkSNARKs usually produce proofs that have constant number of elements.}.
\end{description}

We note that zkSNARKs described in~\cite{DBLP:journals/iacr/GrothM17,DBLP:journals/iacr/BoweG18} present an additional security property that could be desirable, i.e.~
\begin{description}
  \item[Simulation extractability:] $\snark$ is simulation-extractable if from any proof $\pi$ for instance $\inp$ output by an adversary with access to an oracle producing simulated proof on given inputs, it is possible for an extractor to extract a witness $\wit$, such that $(\inp, \wit) \in \REL$.
\end{description}
Obviously, simulation extractability is a stronger property than knowledge soundness. This property is crucial for securing transaction non-malleability (cf.~\cref{sec:transaction-non-malleability}).

\subsubsection{Trusted setup and CRS}\label{sec:crs-generation}
As previously stated, zkSNARKs come with the need to generate a common reference string for the proofs to be computed and verified \cite{DBLP:journals/joc/GoldreichO94}. In the literature it is usually assumed that the CRS comes from a trusted third party that provides the string and effectively disappears (does not take any part in the following computations).
While a trusted party could be named to run the setup, such trust assumption is sometimes unacceptable, especially in the context of public blockchains.
Furthermore, if the CRS was produced maliciously it may compromise the privacy and security of the system breaking zero knowledge and soundness properties~\cite{DBLP:conf/ccs/CampanelliGGN17}.

Such issues raised a special attention around the use of multi-party computation (MPC) ceremonies where a set of users run a sophisticated cryptographic protocol to generate the CRS together.
In that setting, the soundness and zero-knowledge of the system is ensured if at least one participant of the ceremony is honest~\cite{DBLP:conf/sp/Ben-SassonC0TV15,sapling-mpc,DBLP:journals/iacr/BoweGG17}.
More recently, Abdolmaleki et al.~\cite{DBLP:conf/asiacrypt/AbdolmalekiBL017} and Fuchsbauer~\cite{DBLP:conf/pkc/Fuchsbauer18} shown how to achieve zero-knowledge if no party in the MPC protocol is honest.

Recently, Groth et al.~\cite{DBLP:conf/crypto/GrothKMMM18} and Maller et al.~\cite{DBLP:journals/iacr/MallerBKM19} proposed a novel approach for the CRS generation for zkSNARKs that allows to produce a single master CRS $\crsmaster$ for all circuits of given size that can be later securely specialized into another CRS $\crsspec$ for a particular Quadratic Arithmetic Program (QAP).
In other words, the techniques developed by Groth et al., and Maller et al.~allow to reduce the trust required in the trusted setup. Now, one only needs to trust that the master CRS was created honestly, what yields security for all possible $\crsspec$. In our case, one could imagine that we would have a single $\crsmaster$ for all possible mixers and zkSNARK-based protocols, such that each of them use different $\crsspec$ and the security of each of the mixers relies on the generation of the master CRS.

\newcommand{\fs}{\mathsf{FS}}
We note that any kind of non-interactive zero knowledge is impossible in the so-called \emph{standard model} and requires some additional assumptions. As an alternative to the \emph{common reference string model}~\cite{DBLP:conf/crypto/CanettiF01} one could use the \emph{random oracle model}~\cite{DBLP:conf/ccs/BellareR93}, as, e.g.,~in~\cite{DBLP:conf/sp/BunzBBPWM18,DBLP:conf/tcc/Ben-SassonCS16}, and transform any constant-round interactive protocol into a non-interactive one using Fiat-Shamir transformation $\fs$~\cite{DBLP:conf/crypto/FiatS86,DBLP:conf/eurocrypt/PointchevalS96}. However, it is important to note that no efficiently computable function can instantiate the random oracle~\cite{DBLP:conf/stoc/CanettiGH98,DBLP:conf/crypto/Nielsen02}, the security proof for $\fs$ in~\cite{DBLP:conf/eurocrypt/PointchevalS96} works for constant-round protocols only, and it is not certain when the transformation is secure if the random oracle is substituted by a concrete function.

\section{Overview of the protocol}\label{sec:overview-protocol}
The following section dives into the specifics of the \zeth{} protocol. \zeth{} takes its roots in the Zerocash protocol~\cite{zerocash-paper}, but comes with several modifications to work on Ethereum.
Here, we introduce our protocol and give a short example of a possible use case. In the considered scenario we focus on two parties Alice $\alice$ and Bob $\bob$. The parties join $\zeth{}$ and Alice deposits her money into the mixer and makes a secure and private payment to some parties, one of which is Bob.
A variable $\variable{var}$ attributed to a party $\pdv$ is written as $\pdv.\variable{var}$ or simply $\variable{var}$, if $\pdv$ is clear from the context. For the sake of simplicity we assume that each party uses only one of their Ethereum accounts in the mixer. 

\subsection{$\zeth{}$ setup}
In order to make a payment via \zeth{}, several smart contracts need to be deployed on-chain first.
\begin{description}
    \item[$\AddressRegistry$:] (optional) The address registry contract stores the public tuple of the \zethAddresses{}, i.e. encryption keys and public key used to create notes, of all peers.
    \item[$\Mixer$:] The mixer contract handles deposits of funds, private transfers and withdrawals. The contract contains and keeps track of a Merkle tree of commitments (via $\MerkleTree$), the serial numbers, and calls $\Verifier$ (instantiated with the CRS generated from the setup) to verify the provided proof during a private transfer of digital assets.
    \item[$\MerkleTree$:] The Merkle tree contract implements a Merkle tree using $\sha$ as a hash function. The Merkle tree is initialized with zero values at its leaves and is full,  i.e.,~for a tree of depth $d$ it contains $2^{d + 1} - 1$ nodes.
    \item[$\Verifier$:] The verifier contracts contains the CRS generated by the setup and is in charge of verifying proofs provided by payment senders.
\end{description}

The security of \zeth{} relies on the security of the zkSNARK proof system $\snark$, which has to be provided with a CRS. As mentioned in \cref{sec:crs-generation}, we propose to use multi-party computation to generate the CRS using e.g.~\cite{DBLP:conf/sp/Ben-SassonC0TV15,DBLP:journals/iacr/BoweGG17}.
The CRS of the zkSNARK has to be available for various parties of the protocol.
More precisely, the sender of the payment plays the role of a prover $\prover$ and needs the CRS to make an acceptable proof $\pi$.
The proof is checked by a smart contract $\Verifier$, which runs the $\snark$'s verifier routine $\verifier$, thus it has to be provided with the CRS as well.
We emphasize here that the CRS needs to be generated only once for the whole lifespan of the mixer.

\subsection{Joining the mixer}
  To join $\zeth{}$ each party $\pdv$ starts with generating their \zethAddresses{} which are tuples of the form
    \[
    \pdv.\addr \coloneqq (\pdv.\addr_{\pk}, \pdv.\addr_{\sk})\enspace.
    \]
    An address $\pdv.\addr_{\circ}$ consists of $\pdv.a_{\circ}$ and  $\pdv.k_{\circ}$, where $\pdv.a_{\pk}$ is a public address derived from a private $\pdv.a_{\sk}$, $\pdv.k_{\pk}$ is a public encryption key, and $\pdv.k_{\sk}$ is a private decryption key.
    The key pair $(\pdv.k_{\pk}, \pdv.k_{\sk})$ is used to enable the peers to send each other data privately.

    After generating a \zethAddress{}, each party publishes the public tuple $\pdv.\addr_{\pk}$ in $\AddressRegistry$ in order to be able to receive payments from other counterparts.
    Note, this step is not absolutely necessary and, if done, should be done only once for each new joiner.
    It is also feasible for the parties to send each other their public addresses $\addr_{\pk}$ using some other channel of communication (off-chain).
    However, if $\pdv$ wants to transact with, say, $N$ other members of the network they need to send $N$ off-chain messages to inform all of them of their address $\pdv.\addr_{\pk}$. Such a workaround could impede the use of \zeth{}.

\subsection{$\mix$ function}
The core of the mixer consists of a $\mix$ function, which provides a method for secure and private payments between the parties.
$\mix$ covers the main functionalities of $\Mixer$, and provides a unique interface to:
\begin{enumerate}
  \item \label{it:deposit} make a deposit of public funds to the mixer (and \emph{mint} \zethNotes{}),
  \item \label{it:pay} proceed with payments between users via the use of $\Mixer$,
  \item \label{it:withdraw} withdraw funds from $\Mixer$ (redeem public funds).
\end{enumerate}

These three functionalities imply a natural division of $\mix$'s inputs:
\begin{itemize}
  \item The value $v_{\vin}$ which specifies the public value that should be added to the mix. This input is essential for the deposit functionality.
  \item The value $v_{\vout}$ which specifies the public value that should be sent back to the caller. This input is essential for the withdrawal functionality.
  \item And finally, a set of inputs used to represent transfer of funds in $\Mixer$, which contains:
  \begin{itemize}
    \item the commitments $\smallset{\cm^{\new}_i}_{i = 1}^{\coinsOut}$ to the newly generated \zethNotes{} $\smallset{\znote^{\new}_i}_{i = 1}^{\coinsOut}$,
    \item a zero-knowledge proof $\pi$ that assures that
      \begin{enumerate}
        \item the transaction is balanced, i.e.~the sum of the transaction inputs equals the sum of transaction outputs;
        \item if the creation of $\znote^{\new}$ requires spending some old \zethNotes{} $\znote^{\old}$, which commitments are stored in a Merkle tree of root $\rt$, then the user has the right to do so;
      \end{enumerate}
    \item the serial numbers of the spent notes $\smallset{\znote^{\old}_i}_{i = 1}^{\coinsIn}$, to ensure double-spending prevention;
    \item the encryption of the newly created \zethNotes{}, i.e.~ciphertexts $\smallset{\cxx_i}_{i = 1}^{\coinsOut}$ that encrypts $\smallset{\znote^{\new}_i}_{i = 1}^{\coinsOut}$; these ciphertexts are crucial for the recipient of the transfer to be able to spend the notes.
  \end{itemize}
\end{itemize}

Obviously, privacy of deposits and withdrawals is limited since some pieces of information are inevitably revealed as the public balances of the depositor and withdrawer are modified along with $\Mixer$'s balance.
We also note here that the parties \emph{cannot} transfer all of their funds to the mixer since they need their external accounts to pay for the gas necessary to perform $\zeth{}$ operations.
Nevertheless, no one can tell how the deposited value is spread across the newly created \zethNotes{}.
On the other hand, privacy guarantees in $\Mixer$ payments are strong -- nobody can tell which $\znote^{\old}$-s were spent, what is the value of newly created $\znote^{\new}$-s, and who is their recipient.

\subsection{Payments explained practically}
Here we provide an illustrative example of how $\mix$ works.
Now that payer Alice and recipient Bob have generated their \zethAddresses{} (and advertised their public keys), they can start making payments.
As mentioned, we allow Alice to specify up to $\coinsOut$ recipients $\recipient_1, \ldots, \recipient_\coinsOut$.
Assume Alice wants to pay $\recipient_i$ value $v_{\alice \recipient_i}$. Note that one or all of these parties could be Alice herself and all the \zethNotes{} can be zero-valued. Furthermore, commitments of zero-valued \zethNotes{} are indistinguishable from the rest, cf.~\cref{it:commitment}.
One should not be able to tell whether she paid $\coinsOut$ people $\coinsOut$ different amounts, or if she paid $p$ people ($p < \coinsOut$) and took back some change, or just whether she split her funds to herself.

As mentioned, Alice can use two types of funds: her public funds and the funds she controls on the $\Mixer$ contract, i.e.~\zethNotes{} $\znote_\alice$.
Each of them may be used to create new committed notes $\cm^{\new}_1, \ldots, \cm^{\new}_{\coinsOut}$ that will be appended to the Merkle tree of $\Mixer$.
The commitments $\cm^{\old}_1, \ldots, \cm^{\old}_{\coinsIn}$ already contained in $\MerkleTree$ which Alice controls and decides to use in the transaction are called \emph{spent commitments}. The notion of \emph{spent commitment} is technically imprecise, but we use it to ease the understanding of the protocol. In fact, the set of commitments is an ever-growing list, and no one should be able to distinguish between commitments corresponding to notes that have already been used in payments (i.e.~spent commitments), and commitments which associated notes have never been used. Only the user knowing the opening of the commitment can tell whether a commitment was spent or not.

Furthermore, Alice generates a zero-knowledge proof of knowledge $\pi$  that assures she knows the private input $a^{\old}_{\sk, i}$ corresponding to $a^{\old}_{\pk, i}$ used to compute $\cm^{\old}_1, \ldots, \cm^{\old}_{\coinsIn}$, and computes \zethNotes{}' serial numbers $\sn^{\old}_1, \ldots, \sn^{\old}_\coinsIn$.
That assures she is the rightful owner of the funds she tries to spend and no double-spending occurs.
After the proof is computed, Alice calls
\[\mix(\mixinput)\enspace.\]
The values $\mixinput$ constitute the data of a mix transaction $\txmix$.
We describe below, the steps Alice needs to follow in order to transact via $\Mixer$.

\subsubsection{Commitment creation}\label{it:commitment}
Alice creates $\coinsOut$ commitments $\cm_1^{\new}, \ldots, \cm_{\coinsOut}^{\new}$ to $\coinsOut$ notes of values (not necessarily distinct) $v_1, \ldots, v_\coinsOut$.
For each of the newly created commitment Alice picks randomness $r^{\new}_i, s^{\new}_i, \rho^{\new}_i$ and computes
\[
    \cm^{\new}_i = \com_{s^{\new}_i} (v_i \| \com_{r^{\new}_i} (a^{\new}_{\pk, i} \| \rho^{\new}_i))\enspace.
\]
With the commitments created, Alice proceeds to the next step.

\subsubsection{Preparation of ciphertexts}

Zeth notes $\znote$ are defined as openings of the commitments $\cm$. More precisely, for a  commitment $\cm^{\new}_i$, we define
\[
    \znote_i = (a^{\new}_{\pk, i}, v_i, \rho^{\new}_i, r^{\new}_i, s^{\new}_i)
\]
and denote by $\cxx_i$ the encryption of the note, $\cxx_i = \enc_{\recipient_i.k_{\pk}} (\znote_i)$.
Note that, if Alice wants to transfer some notes back to her (e.g.~as a change), she encrypts these notes using her own encryption key $\alice.k_{\pk}$. Since the encryption scheme is key-private, no one can tell which key was used to encrypt the message. Thus, it effectively hides the recipient. Moreover, the ability of the $\Mixer$ contract to broadcast $\coinsOut$ ciphertexts in a single smart contract call enables Alice to pay up to $\coinsOut$ recipients in the same transaction and paves the way for more complex payment schemes.

\subsubsection{Zero-knowledge proof generation}\label{sec:zkstatement}
Since commitments $\cm^{\new}_i$ may be associated to some value and spent in the future, Alice needs to provide a proof that these commitments are well-formed and that she had enough funds to create the associated notes.
    To that end, Alice picks her $\coinsIn$ commitments $\cm^{\old}_1, \ldots, \cm^{\old}_\coinsIn$ from $\MerkleTree$ and shows in zero-knowledge that given the primary inputs (the Merkle tree's root $\rt$, serial numbers $\sn^{\old}_1, \ldots, \sn^{\old}_\coinsIn$, commitments $\cm^{\new}_1, \ldots, \cm^{\new}_{\coinsOut}$, $v_{\vin}$, and
    $v_{\vout}$), and the auxiliary inputs (input commitments addresses in the Merkle tree $\cmAddr^{\old}_1, \ldots, \allowbreak \cmAddr^{\old}_\coinsIn$, input notes $\znote^{\old}_1, \ldots, \znote^{\old}_\coinsIn$, Merkle paths to input commitments $\mkPath^{\old}_1, \allowbreak \ldots, \allowbreak \mkPath^{\old}_\coinsIn$, sender secret keys $a^{\old}_{\sk, 1}, \ldots, a^{\old}_{\sk, \coinsIn}$):
\begin{enumerate}
    \item For each $i \in \range{1}{\coinsOut}$,
    \begin{enumerate}
        \item $\cm^{\new}_i = \com_{s^{\new}_i}(v^{\new}_i \| k^{\new}_i)$,
        \item $k^{\new}_i = \com_{r^{\new}_i}(a_{\pk, i}^{\new} \| \rho^{\new}_i)$.
    \end{enumerate}
    \item For all $i \in \range{1}{\coinsIn}$, commitment $\cm^{\old}_i$
    \begin{enumerate}
    \item appears in the Merkle tree of root $\rt$ at address $\cmAddr^{\old}_i$ and $\mkPath^{\old}_i$ is a valid Merkle path to this commitment,
    \item the commitment used in the payment is correctly formed, i.e.:
    \begin{itemize}
        \item $\cm^{\old}_i = \com_{s^{\old}_i}(v^{\old}_i \| k^{\old}_i)$,
        \item $k^{\old}_i = \com_{r^{\old}_i}(a_{\pk, i}^{\old} \| \rho^{\old}_i)$,
        \item $a_{\pk}^{\old, i} = \prf^{\addr}_{a^{\old}_{\sk, i}} (0)$,
        \item $\sn^{\old}_i = \prf^{\sn}_{a^{\old}_{\sk, i}} (\rho^{\old}_i)$, and
        \item the condition $v^{\old}_i \cdot (1 - \enforce) = 0$ is satisfied. The boolean value $\enforce$ is related to an equality check between the given Merkle root $\rt$ and the Merkle root $\rt'$ resulting from the check of the Merkle path $\mkPath^{\old}_i$. More precisely, if $v^{\old}_i > 0$, then $\enforce = 1$ to make sure that the commitment of the associated note is in the Merkle tree, and that $\rt = \rt'$. Nevertheless, if $v^{\old}_i = 0$, then $\enforce$ could be $0$, which represents the fact that the equality of the roots is not required\footnote{The equality check between two bit strings
        $b_1$ and $b_2$ can be carried out by a succession of constraints in the form $\enforce \cdot (b_1[i] - b_2[i]) = 0$ for each bit of the bit strings, where $b_1[i]$ ($b_2[i]$) represent the $i$-th bit of $b_1$ ($b_2$, resp.). Obviously, if $\enforce = 1$, then the constrain is satisfied only if the bit strings are equal. On the other hand, for $\enforce = 0$ the constraint is always satisfied.}.
        In other words, this extra condition enables to support zero-valued (or dummy) \zethNotes{}.
    \end{itemize}
    \end{enumerate}
\item The joinsplit equation below holds, and no value is created by the transaction:
\begin{equation}
    \label{eq:joinsplit}
    v_{\vin} + \sum_{i = 1}^{\coinsIn} \znote^{\old}_{i}.v =  \sum_{i = 1}^{\coinsOut} \znote^{\new}_i.v + v_{\vout}\enspace,
\end{equation}
where $v_{\vin}, v_{\vout}$ are the public input value and public output value resp, and where $\znote_{i}.v$ denotes the value of the note $\znote_{i}$.
\end{enumerate}

    In the rest of this paper, we denote by $\RELCIRC$ the NP-relation that consists of all pairs of primary input $\inp$ and auxiliary input $\wit$ that satisfy the constraints above.
    Note, that the statement as presented is adapted from \project{Zcash\,Sprout}~\cite{zcash-protocol-spec}, and is a bit different than the one in the original Zerocash paper.
    However, the computation of $\sn$ and $\cm$ follows closely the construct detailed in~\cite{zerocash-paper}, which asks for some additional security mechanisms around the choice and the use of $\rho$ and $r$ to avoid the creation of malicious notes that cannot be spent\footnote{According to~\cite{zerocash-paper}, each serial number is only computed from the first $254$ bits of $\rho$. As a consequence, it is possible for a malicious payer to create four notes with $\rho_0, \ldots, \rho_3$ such that $\smallset{\rho_i}_{i = 0}^{3}$ only differ on the last two bits. In such case, the four generated notes are valid but share the same serial number $\sn$. Thus, after the first note of the four is spent, it is impossible to spend the rest. An easy way to solve this problem is to use all the bits of $\rho$ in the derivation of $\sn$, for instance.}.
    Last but not least, the above-mentioned statement does not contain components that were added in~\cite{zerocash-paper} to provide transaction non-malleability. To achieve this property we propose to either use a non-malleable zkSNARK like \cite{DBLP:journals/iacr/GrothM17,DBLP:journals/iacr/BoweG18} along with an additional check used to bind the ciphertexts computation to the rest of the statement, or add the additional elements as specified in \cite{zerocash-paper} and \cite[Section 4.15.1]{zcash-protocol-spec} and in the Zcash code base. We use such an extended statement in~\cref{sec:transaction-non-malleability}.

    \subsubsection{$\mix$ function call}
    After generating the commitments $\smallvec{\cm^{\new}_i}_{i = 1}^{\coinsOut}$, computing the serial numbers $\smallvec{\sn^{\old}_{i}}_{i = 1}^{\coinsIn}$ and the proof $\pi$, and encrypting the notes $\smallvec{\cxx_i}_{i = 1}^{\coinsOut}$, Alice is ready to call
    \[
      \Mixer.\mix(\mixinput)\enspace.
    \]
    The $\Mixer$ contract passes all $\mix$'s inputs, except $\smallvec{\cxx_i}_{i = 1}^{\coinsOut}$, to the $\Verifier$ contract to run the zkSNARK verification algorithm $\snark.\verifier$. Finally, $\Mixer$ accepts iff:
    \begin{itemize}
      \item the given Merkle root $\rt$ is valid (i.e.~corresponds to one of the roots that has fingerprinted the Merkle tree of commitments maintained by the smart contract),
      \item none of the serial numbers $\sn \in \smallset{\sn^{\old}_{i}}_{i = 1}^{\coinsIn}$ is in the list of serial numbers that have already been used,
      \item the zkSNARK verifier $\snark.\verifier$ accepts the proof $\pi$,
      \item the sender's balance is at least $v_{\vin}$.
    \end{itemize}
    In that case, $\Mixer$ appends $\smallset{\sn^{\old}_{i}}_{i = 1}^{\coinsIn}$ to the list of serial numbers, adds $\smallset{\cm^{\new}_i}_{i = 1}^{\coinsOut}$ as leaves to \comment{the }$\MerkleTree$; and broadcasts $\smallset{\cxx_i}_{i = 1}^{\coinsOut}$. Otherwise, it rejects.
	One should note that it is not possible to link the commitments in the Merkle tree with their associated publicly disclosed serial numbers. This holds since, the serial numbers are derived from private data only known by the owner of the notes associated with the commitments.
	This also ensures that a user appending a commitment in the tree, as part of a payment to another user, does not know the serial number in advance and thus, cannot tell when the recipient spends the received funds.
  Moreover, given a note $\znote$, its associated commitment $\cm$ and the given serial number $\sn$, it is intractable to compute a new serial number for $\znote$ different from $\sn$. Thus, disclosing the serial numbers publicly allows parties to detect and reject attempts of double spending of the same note.

    \subsubsection{Payment reception}\label{sec:payment_reception}
    Assume $v_{\alice \bob}$ is a value that Alice agrees to sent Bob.
    To carry out such payment she computes all necessary inputs to the $\mix$ function and executes it. Although no more is required from Alice, Bob cannot just accept that $\mix$ function was executed properly to recognize she fulfilled her obligations.
    For example, cheating Alice could transfer less money than $v_{\alice \bob}$. Since the $\Mixer$ contract does not have access to the value Alice and Bob agreed to exchange, it only checks constraints to ensure that the system remains sounds, and thus, do not compare Alice's committed value $v$ with $v_{\alice \bob}$. To make sure that the payment was done as supposed, Bob has to perform some additional actions.

    First of all, Bob needs to fetch the transferred notes. To that end he observes the events emitted by Alice's call to the smart contract and tries to decrypt all broadcast ciphertexts using his decryption key $\bob.k_{\sk}$.
    Say that among the $\coinsOut$ ciphertexts $\coinsOut'$ decryptions were successful, resulting in a set of new \zethNotes{} $\smallset{\znote_{j}}_{j = 1}^{\coinsOut'}$ owned and controlled by Bob. (Note that Bob does not need to know $\coinsOut'$ in advance.)
    This operation is done by calling a function $\receive$ that scans $\Mixer$'s events and tries to decrypt all ciphertexts that was posted after Bob's last call to the function. Every successful event decryption provides Bob with a new note that is then stored in his wallet.

    However, even though Bob received the notes, he needs to check that he can use them and that their values sum up to $v_{\alice \bob}$. To that end, Bob recomputes all commitments $\smallset{\widetilde{\cm}^{\new}_{j}}_{j = 1}^{\coinsOut'}$ given corresponding $\smallset{\znote_j}_{j = 1}^{\coinsOut'}$. Then, he checks if:
    \begin{itemize}
    	\item all commitments $\smallset{\widetilde{\cm}^{\new}_{j}}_{j = 1}^{\coinsOut'}$ are among commitments $\smallset{{\cm}^{\new}_{i}}_{i = 1}^{\coinsOut}$ just appended to the Merkle tree in $\MerkleTree$ by Alice;
    	\item he is able to compute all the serial numbers associated with commitments  $\smallset{\widetilde{\cm}^{\new}_{j}}_{j = 1}^{\coinsOut'}$ and that they have not been published yet;
    	\item the values associated with the commitments $\smallset{\widetilde{\cm}^{\new}_{j}}_{j = 1}^{\coinsOut'}$ sum up to $v_{\alice \bob}$;
    \end{itemize}
	and accepts the payment if they all hold.
  Now Bob can follow the same steps Alice did to use the notes he received and pass them on.

\section{Abuse mitigation and best practices}\label{sec:abuse-mitigation}\label{sec:best-practices}
It is important to note that even the best private payment system cannot provide much privacy if its users, who are supposed to be protected and anonymous, do not follow some best practices, misuse the system, or do not know its limitations. Even Zcash users' anonymity can be breached if they behave improperly~\cite{DBLP:conf/uss/KapposYMM18}.
\cref{sec:overview-protocol,sec:security-guarantees} give an overview of how the \zeth{} protocol works and the security model it follows.
This section presents potential ways to exploit the system to infer valuable information about the ongoing payments showing the limits of the security model we use and presents a few best practices to mitigate attack vectors as much as possible.

	\paragraph{Keep the funds in the mixer as long as possible}
    As explained in \cref{sec:zeth-protocol}, the state maintained by the Ethereum blockchain leaks some information about users of the system.
    As a consequence, the longer users make payments via the mixer, without transfers from/to Ethereum accounts, the better. This leaks less data to passive attackers who monitor the account balances of users of the mixing service.

	\paragraph{Beware of almost empty $\MerkleTree$}
    As one might have noticed in \cref{sec:overview-protocol}, if the Merkle tree of commitments $\MerkleTree$ is almost empty when Alice and Bob want to make a secure payment, then their privacy is at risk as they could be identified as the sender and recipient of the payment.
    Assume a scenario when there is almost no commitments in $\MerkleTree$, then some funds are added by Alice and some funds are withdrawn by Bob. In some circumstances, an adversary who observes the contract can guess that Alice made payment to Bob and the payment was at least of value withdrawn by the latter.
    Thus, it is important to avoid withdrawals until the number of non-zero leaves (commitments appended) of \comment{the }$\MerkleTree$ is of reasonable size.

    On the other hand, a large $\MerkleTree$ cannot be considered absolutely safe. Since the adversary can easily add many $\MerkleTree$ entries by doing transactions between accounts controlled by themselves, Alice may be tricked to think $\MerkleTree$ is big enough, while in fact all entries but Alice's are controlled by a single eavesdropping party.

    \paragraph{Small user anonymity set}
    Similar situation occurs when there is very few active parties using the $\Mixer$ contract. If Alice and Bob are the only parties using $\Mixer$, then an observant adversary can assume that each payment made by Alice has Bob as a recipient. Furthermore, if Bob withdraws funds from the contract despite he has not made any deposit, the adversary could conclude that the funds were transferred by Alice.

    We suppose here that we have a public registry of users of the $\Mixer$ contract (as e.g.~$\AddressRegistry$).
        It is important to understand that even though a large set of users of $\Mixer$ often comes with the belief of stronger privacy guarantees this is not always the case. In fact, similarly to the case of a small set of commitments in $\MerkleTree$, an adversary could easily register a lot of parties in the $\AddressRegistry$ contract by paying only a small fee. In that case, Alice could have the impression that the registry contains a lot of entries, while almost all of them are controlled by a single entity, leading to a system with insufficient privacy guarantees~\cite{DBLP:conf/iptps/Douceur02}\footnote{We note here that the cost of such an attack is very low. Adding an entry to the on-chain address registry $\AddressRegistry$ is very cheap and basically consists in paying the gas required to add an entry to the storage of $\AddressRegistry$, along with the intrinsic gas cost of an Ethereum transaction.}.

    \paragraph{Split funds in the mixer}
    It is a good practice for users to make payments to themselves to spread their wealth across a set of commitments and obfuscate their balance. Note that these payments may be dummy payments of value $0$, which are indistinguishable from non-zero payments and introduce the necessary noise in the system.
    However, such operations come at the cost of paying for gas for the execution of the $\mix$ function.

    \paragraph{Use the most recent Merkle root to enhance the mixing}
    As the set of all commitments is stored in a Merkle tree, it is fingerprinted by the tree's root. Hence, whenever a new commitment is appended as a leaf of the tree, the value of the root changes.
    Importantly, if a user uses an older version of the Merkle tree to spend their funds, the associated proof is still valid. However, such behaviour is not advised.
    As the tree is ever-growing, the number of non-empty leaves at time $t_0$, denoted by $n_{t_0}$, is smaller or equal to this number at time $t_1$, denoted by $n_{t_1}$, for $t_0 < t_1$.
    Assume a transfer is made at time $t_1$ using the root of the tree from $t_0$. Then, even though the tree contains $n_{t_1}$ leaves an adversary can tell that the commitments used in the transaction are among those that were appended to the tree before $t_0$ what limits anonymity.

\paragraph{If decryption of zeth notes fails}
Assume that Alice and Bob have a contractual agreement and that Alice owes $k$\,\ETH{} to Bob in exchange for some goods.
Note that when Alice transfers funds to Bob she is supposed to encrypt a number of \zethNotes{} of agreed values using Bob's encryption key. Moreover, Bob can pass the funds on only if he decrypts them.
If Alice wishes to trick Bob, she could decide to encrypt her \zethNote{} using her own encryption key. That way, instead of making a payment to Bob, she makes a payment to herself and waits for Bob to send her goods.
Nevertheless, this issue could be solved by requiring Bob to withhold sending Alice's goods until he manages to decrypt the notes that match with the expected value for the goods\footnote{Basically, Bob withholds sending Alice's goods until he successfully ran the $\receive$ function.}. Unfortunately, now Bob can behave maliciously and accept the payment without sending the goods. This problem, referred to as \emph{contingent payment} is out of the scope of this paper.

In case Bob receives a payment that is not correct or just fails to receive any payment from Alice, we propose to use a dispute mechanism coupled with a reputation system (again, out of scope of this paper) which could be a way to stop malicious senders.
In such scenario, Bob could make a zero-knowledge proof that no ciphertext broadcast by $\Mixer$ matched the contract he had with Alice.
In that case, Alice's reputation on the system could be severely damaged and may result in fewer users willing to transact with her.

\paragraph{$\Mixer$ users distinguishability}

An important drawback of \zeth{} comes from the fact that users of $\Mixer$ are distinguishable. That is, it is feasible for an adversary to tell the difference between users using the mixer and those who do not, i.e.~who have never used $\AddressRegistry$ or called the mixing contract.
Hence, it is important to be aware that a party registering an address to $\AddressRegistry$ discloses that they perform transactions via \zeth{} or help other parties to do that.

However, the recipient of a \zeth{} payment can remain unrecognized as a user of the system as long as they transmitted their $\addr_{\pk}$ to the sender using off-chain communication; and as long as they do not manipulate the funds they control on the mixer.
A party is identified as a user of \zeth{} as soon as they make a call to the mixer or any other \zeth{} smart contract. Leaking such information may be benign in some cases, but sometimes it may put the user at risk, e.g.~if the user lives in a country ruled by an authoritarian regime that forbids using privacy-enhancing technologies.


\paragraph{Failures and backup}
The design of \zeth{} presents some desirable properties related to hardware failure or loss of the \zethNotes{}.
In fact, provided that a user has a backup of their \zethAddresses{}, they can easily retrieve the set of their \zethNotes{} by scanning the chain in search for events emitted by the mixer contract.
As described earlier, $\Mixer$ broadcast encrypted \zethNotes{}, thus a user who downloads and tries to decrypt all emitted ciphertexts, eventually retrieves the set of all \zethNotes{} they should control.

\paragraph{Forward secrecy}
Using a single \zethAddress{} for all incoming payments puts a user at risk.
An adversary who retrieves user's decrypting key $k_{\sk}$ at time $t$ could expose all communications that happened prior to this time and thus violate forward secrecy.
A way to mitigate this, would be to renew \zethAddresses{} periodically to ensure that \zethNotes{} are encrypted under different encryption keys over time.

\paragraph{The threat of a quantum attacker}
Using encrypted broadcast implies storing ciphertexts on a publicly available chain.
A direct consequence is that they are subject to quantum attacks, which violates the privacy promises of the protocol, unless the ciphertexts are encrypted using post-quantum secure encryption scheme.

\comment{
\subsection{Blockchain size}
Being able to leverage publicly verifiable zkSNARKs proofs constitutes a building block of our protocol.
Whilst on-chain verification of zkSNARKs proof is fundamental to keep a consistent state under consensus, it also comes at the price of storing every proof on-chain. Despite the fact that zkSNARKs proofs are short, storing all proofs on chain adds further load on the miner nodes that would need to increase their storage capabilities if the rate of transactions going through our mixing contract was to become significant.

[TODO: Nuance this with the ability we have to proceed to $m$ payments in a single call to the smart contract]
}

\section{\zeth{} for permissioned chains}\label{sec:public-vs-permissionned-chains}

In this section we detail a few properties of \zeth{} regarding its use in consortium and permissioned chains. We note that threat models between permissioned and public chains differ in many ways. For example, permissioned networks offer control over the parties that want to join, can efficiently counter Sybil attacks and if parties behave maliciously, they can be excluded from the chain and never allowed to join again.

\paragraph{CRS generation}
As mentioned in \cref{sec:overview-snarks}, a trusted setup needs to be carried out once for the protocol to work. While delegating the generation of the CRS to a trusted party inside or outside the system could be acceptable in some circumstances, such trust assumptions are unacceptable in most cases.

In order to limit the trust put in the party in charge of generating the CRS, multi-party protocols could be considered.
However, it is still required for the users of the system to believe that among all the parties that are involved in the CRS generation at least one is honest.
In the other case, colluding parties could prove false statements and, effectively, create money ``out of thin air".

By its very nature a public blockchain project is opened to everyone. As a consequence, the set of potential users of the system can theoretically contain all humans living on Earth.
Although often all system users are welcomed to join multi-party CRS generation, practice shows that it is not usually the case. One example of this observation is the recent Sapling MPC implemented by Zcash. Despite the fact that the procedure was opened to everyone, only around $90$ people\footnote{The list can be found on \url{https://github.com/zcash-hackworks/sapling-mpc/wiki}.} joined the MPC ceremony in order to generate the CRS.
Furthermore, in a public blockchain there is no way of telling how many of these $90$ people were in fact different, who they represent, or what are their incentives. That is, a dishonest user could simply generate and register multiple identities hoping for gaining unfair impact on the final shape of the generated CRS.

Nevertheless, in permissioned blockchains the situation is quite different and it is justified to assume that all the network members are known. Moreover, some requirements can be enforced to ensure that all parties involved in the network participate in the distributed trusted setup generation. For example, access to the network may only be granted to the parties who have participated in the CRS generation.

Furthermore, one can assume that the parties also \emph{do not want to} pass on the CRS generation.
This is since members of consortium chains are often actors that practice and evolve in very similar economic sectors. Hence, every member has an incentive to acquire information about the activity of the others. In fact, being able to gather intelligence about other members or being able to violate the soundness of the system could lead to a competitive advantage. Thus, it is reasonable to assume that all members of the network are incentivized to take part in the generation of the CRS, rather than just use the trusted setup provided by their competitors. As a consequence, it is justified to remove the threat of full collusion regarding the generation of the CRS from the threat model.

\paragraph{Privacy leakages}
As mentioned, Ethereum addresses of parties using the mixer contract are revealed since they all need to pay gas to call the mixer's function.
While such leakage might be problematic in public chains, this is not necessarily an issue in permissioned networks. In fact, it could be expected that the \zethAddress{} of each network member is added to $\AddressRegistry$ as part of the pemissioning mechanism for instance.

\paragraph{A note on Sybil attacks and distinguishability}
As noted in \cref{sec:best-practices}, a small number of registered players in $\AddressRegistry$ may allow an adversary to efficiently connect senders and recipients of payments. Furthermore, an adversary can fool honest parties to believe that there are enough parties in the mixer to provide necessary security, while that is not the case, cf.~\cref{sec:abuse-mitigation}.
Fortunately, in a permissioned blockchain Sybil attacks can efficiently be prevented, since joining the chain is not open to everyone and one can assume that each party therein is somehow verified (e.g.~when the system remains compliant with the Know Your Customer (KYC) requirements).
This property allows for more precise measurements of privacy level of the mixer.
For example, given access to $\AddressRegistry$ one can credibly estimate the number of all involved parties and judge for themselves whether this number is satisfiable for them or not.

\section{Implementation}\label{sec:implementation}

\begin{figure}
  \centering
  \includegraphics[width=1\textwidth]{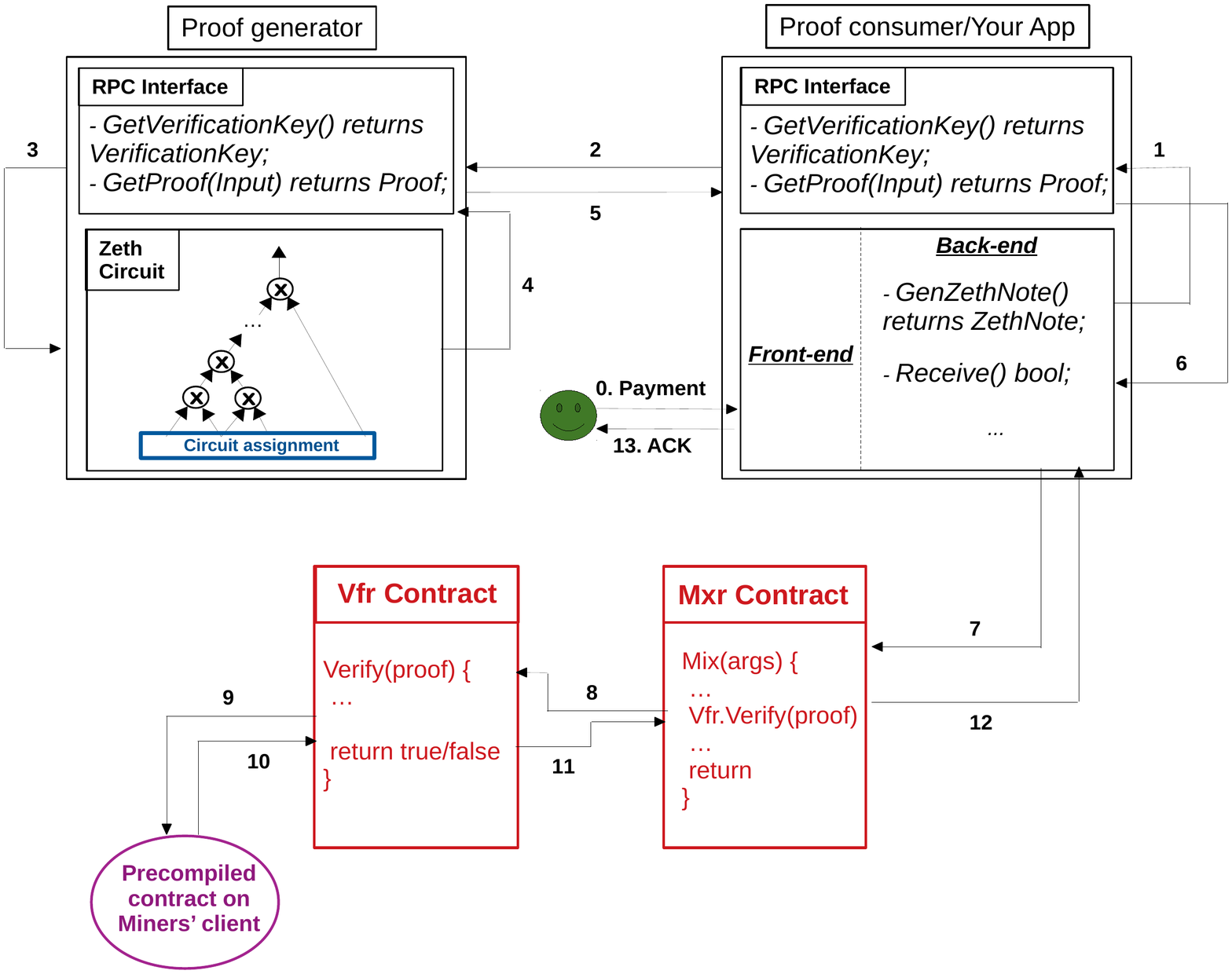}
  \caption{An overview of the architecture and flow of function calls for the current proof of concept.}
  \label{fig:architecture}
\end{figure}

In this section we describe the proposed proof of concept implementation of \zeth{}\footnote{\zethgithub}.
Our implementation relies on the \project{libsnark}\footnote{\url{https://github.com/scipr-lab/libsnark}} library and makes use of the precompiled contracts for the elliptic curve operations on \texttt{bn256} introduced in Ethereum after the Byzantium hardfork.
In order to implement \zeth{}, we follow the \emph{separation of concerns}~\cite{Dijkstra1982} approach. Hence, the proposed architecture is modular and comports different components, each of which has its own role. The components communicate with each other using Remote Procedure Calls (RPC), which are implemented using the \project{gRPC} framework\footnote{\url{https://grpc.io/}} along with \project{Protocol\,Buffers}\footnote{\url{https://developers.google.com/protocol-buffers/}}. The overall architecture of the proof of concept is presented in~\cref{fig:architecture}.

\subsection{Overall architecture}

\subsubsection{Proof generator ($\proofGenerator$)}
The core component of \zeth{} is \emph{Proof Generator} ($\proofGenerator$).
This component holds a Rank 1 Constraint System (R1CS) representation for the circuit corresponding to relation $\RELCIRC$\footnote{The circuit was taken from the Zcash code base (\url{https://github.com/zcash/zcash/tree/master/src/zcash/circuit}) and adapted to meet our needs.}, cf.~\cref{sec:zkstatement}, and presents an RPC interface to receive and answer requests from \emph{Proof Consumers} ($\proofConsumer$).

R1CS is an $\npol$-com\-ple\-te relation of tuples $((\AA, \BB, \CC, \vec{v}), \vec{w})$, where $\AA, \BB, \CC$ are matrices and $\vec{v}, \vec{w}$ are vectors over some finite field $\FF$.
The relation is satisfied iff $\AA (1, \vec{v}, \vec{w})  \circ \BB (1, \vec{v}, \vec{w}) = \CC (1, \vec{v}, \vec{w})$.
$\AA, \BB, \CC$ represent the circuit gates (left and right inputs, and an output), $\vec{v}$ and $\vec{w}$ are public and private inputs respectively~\cite{ZKProofImplementationTrackProceeding,ZKProofSecurityTrackProceeding}.
This representation was prepared using the \project{libsnark} library.

In our PoC starting $\proofGenerator$ triggers the generation of a CRS for the zkSNARK for $\RELCIRC$
and results in proving and verification keys, denoted by $\crs_{\prover}, \crs_{\verifier}$ resp.
The keys are stored on disk and the server starts waiting for incoming requests.
As of now, the proposed RPC interface is limited to only two types of calls, i.e~$\getVerificationKey$ and $\getProof$. The first call reads the verification key $\crs_{\verifier}$ and sends it over to the caller. The second, $\getProof$, takes an input provided by $\proofConsumer$, i.e.~an assignment to the witness variables, generates the proof, and returns it to the caller.

To ease the use of \zeth{}, $\proofGenerator$ is run in a Docker container\footnote{\url{https://www.docker.com/}} listening on a given port. This container could be launched on a user's machine in the background as a daemon process waiting to receive requests to generate proofs.
Since proof generation is computationally expensive, $\proofGenerator$ is written in \project{C++} for the sake of efficiency.

\subsubsection{Proof consumer ($\proofConsumer$)}
The proof consumer is a module responsible for requesting proofs to $\proofGenerator$.
This component gets the instance and its witness as input. It does not hold an R1CS representation for the statement or carry out any operation to generate the proofs (contrary to the proof generator). In fact, the $\proofConsumer$ does not even need to know the CRS required to prepare or verify the proof.

Since no expensive computation is expected to be carried out on $\proofConsumer$, it is reasonable to use any high-level programming language without worrying too much about the performance overhead that can be introduced.
In its current state, the proof of concept does not provide any command line interface (CLI) nor front-end to ease payments via $\Mixer$ and improve user experience. Development of applications to ease user experience is left as further work and improvements.

\subsubsection{Separate CRS generator}
It is important to highlight that centralizing the generation of the CRS in the component in charge of generating proofs comes with very high trust assumptions and could translate in a violation of the soundness of the system.
More precisely, if $\proofGenerator$ generates the CRS, it knows the CRS's trapdoor what allows the production of fake, yet acceptable, proofs by running the simulator algorithm of the underlying zkSNARK.
We emphasize here that the secure CRS generation is out of the scope of the presented proof of concept.

A natural improvement of our proof of concept would be to introduce a separate~\emph{CRS Generator} ($\CRSGenerator$) component in charge of generating and publishing the CRS.
$\CRSGenerator$ could either be a single trusted component or one of many components involved in an MPC ceremony that generates the CRS.
Choosing the method to generate the CRS depends on the trust assumptions of the system. We recommend using an MPC ceremony.

Adding the CRS generator as a new component in the architecture comes with a few changes. First of all, new RPC interfaces need to be introduced in order to make other components able to fetch the proving and verification keys from $\CRSGenerator$.
The interfaces are required, since $\proofGenerator$ needs to have access to $\crs_\prover$ to be able to generate a proof. Likewise, the $\Verifier$ smart contract needs to have access to the verification key $\crs_\verifier$ to be able to run the verification routine of the underlying zkSNARK.
Moreover, it should also be possible to require $\CRSGenerator$ to publish the generated CRS $(\crs_\prover, \crs_\verifier)$ to make it easily accessible to users of the system.
While the CRS could be published on a web-page, it is necessary to keep in mind that the server storing and serving the CRS needs to be assumed non-malicious and always online, otherwise it could be possible to change the CRS and advertise a malicious one. To that end, keeping the generated CRS on-chain as part of a smart contract's storage is appealing. However, the size of the proving key is often too big for this solution to be considered.
We let the implementation details of $\CRSGenerator$, along with the design changes it implies, as further work to the interested reader.

\subsubsection{The mixer smart contract}
While both $\proofGenerator$ and $\proofConsumer$ run on a user's machine, $\Mixer$ is a smart contract, and thus gets compiled down to an EVM bytecode and executed on-chain.
This smart contract maintains the Merkle tree of commitments, the list of serial numbers that have been revealed as payments were made, and is in charge of calling the zkSNARK verification algorithm along with some additional checks to keep the system sound. We note that the zkSNARK used in the system is publicly verifiable and the verification is efficient enough to be carried out on-chain.
The core function of $\Mixer$ is the $\mix$ function described on~\cref{fig:mixcode}.

\begin{figure}
	\centering
	\fbox{
		\procedure[linenumbering, syntaxhighlight=auto, addkeywords={abort}]{$\Mixer.\mix(\mixinput)$}{
			\validProof \gets false \\
			\pcif (\neg \MerkleRootList.\contain(\rt)) \pcthen\ abort\ \pcendif\\
			\pcfor i \in \range{0}{\coinsIn} \pcdo \\
			\pcind \pcif (\SerialNumberList.\contain(\sn_{i})) \pcthen\ abort \\
			\pcind \pcelse  \SerialNumberList.\insertFunc(\sn_{i})\ \pcendif \\
			\pcendfor \\
			\validProof \gets \Verifier.\verify(\verifyinput) \\
			\pcif (\neg \validProof) \pcthen\ abort\ \pcendif \\
			\pcif (\neg \isEqual(v_{\vin}, \EthTxObjValue)) \pcthen\ abort\ \pcendif \\
			\pcfor i \in \range{0}{\coinsOut} \pcdo \\
			\pcind \MerkleTreeVar.\insertLeaf(\cm^{\new}_{i}) \\
			\pcendfor \\
			\pcif(v_{\vout} > 0) \pcthen\ \sendValue(v_{\vout}, \EthTxObjSender)\ \pcendif \\
			\rt' \gets \MerkleTreeVar.\getMerkleRoot() \\
			\MerkleRootList.\insertFunc(\rt') \\
			\pcfor i \in \range{0}{\coinsOut} \pcdo \\
			\pcind \broadcastCiphertext(\cxx_{i}) \\
			\pcendfor \\
			\broadcastMerkleRoot(\rt') \pccomment{(optional): Broadcast the \text{new} Merkle root}\\
			\pcreturn
		}
	}
	\caption{Pseudocode of the $\mix$ function of $\Mixer$. Here, $\mathsf{EthTxObj}$ denotes the Ethereum transaction object. It is worth remembering that the execution of a smart contract function is atomic. Thus, if an \textbf{abort} instruction is reached, all the modifications made to the state in the previous instructions are rolled back.}
	\label{fig:mixcode}
\end{figure}

\subsubsection{Paying with \zeth{} -- flow of function calls}
Given the main components of the proof of concept, this section describes the flow of function calls during a \zeth{} payment.
The section follows and explains the steps described on~\cref{fig:architecture}.
\begin{description}
  \item [Step 0:]
      A payment is triggered by a user of the system, who also specifies the recipient(s) and value(s). The selection of the notes to spend can either be detailed by the user, or managed by software.
  \item [Steps 1 and 2:]
    \begin{inparaenum}[(1)]
      \setcounter{enumi}{0}
      \item Randomness is generated for the newly created \zethNotes{} which are then encrypted using the recipients' encryption keys.
      \item An RPC call is made to $\proofGenerator$ to generate a proof. The call passes a witness for the instance being proven.
    \end{inparaenum}
  \item [Steps 3 to 5:]
    \begin{inparaenum}[(1)]
      \setcounter{enumi}{2}
      \item The RPC call is executed and
      \item the proof is generated from the given inputs.
      \item The obtained proof is sent back to the client.
    \end{inparaenum}
  \item [Steps 6 and 7:]
  \begin{inparaenum}[(1)]
    \setcounter{enumi}{5}
    \item The proof is received and
    \item used as an argument, along with the encrypted \zethNotes{}, to the $\mix$ function.
  \end{inparaenum}
  \item [Steps 8 to 11:]
  \begin{inparaenum}[(1)]
    \setcounter{enumi}{7}
    \item $\mix$ is executed and the on-chain verification of the zkSNARK proof is done via the $\Verifier$ contract.
    \item[(9 - 10)] The contract leverages the precompiled contracts for elliptic curve operations on $\texttt{bn256}$ written in the miner's Ethereum client.
    \item[(11)] $\Verifier$ contract returns \texttt{true} if the provided proof is accepted and \texttt{false} if that is not the case.
  \end{inparaenum}
  \item [Step 12:]
    The events triggered by the execution of $\mix$ are emitted.
  \item [Step 13:]
    The $\receive$ function runs in the background on the client. It listens and tries to decrypt all emitted ciphertexts. If a decryption succeeds, a payment is received and the user interface is updated to reflect the reception of a payment.
\end{description}

\subsection{Remarks}

\subsubsection{Deploying $\proofGenerator$ and $\proofConsumer$}
The architecture described above uses different components communicating via RPC calls. Every component could easily be deployed on a separate machine without affecting the way \zeth{} works. However, the threat model associated with the presented proof of concept only considers the case where the proof generator and proof consumer run on trusted machines that are physically accessible, owned and controlled by the user, and that communicate via a secure channel.
In the proposed PoC, the machine running $\proofGenerator$ is represented by a Docker container, whereas $\proofConsumer$ runs directly on the user's machine (i.e.~not in a container). Both processes communicate using a virtual loopback interface. As a consequence, the traffic generated by the communicating processes $\proofGenerator$ and $\proofConsumer$ does not leave the user's machine. Hence why we consider this communication channel to be secure.

Deploying each component on a different and potentially untrusted machine is facilitated by the modularity of the proposed architecture, and can easily be done with very few modifications to the code. Nevertheless, we emphasize that the threat model needs to be modified accordingly depending on where each component runs, who owns and controls the pieces of hardware running $\proofGenerator$ and $\proofConsumer$, how data (keys, \zethNotes{}) are managed and accessed, and so forth. It is highly probable that additional integrity and security checks would need to be implemented to keep the system secure.

\subsubsection{Changing the programming languages}
In the proposed proof of concept $\proofGenerator$ and $\proofConsumer$ communicate via RPC calls and are written in different programming languages.
Since $\proofGenerator$ runs expensive operations related to the CRS generation and proof computation it is advised to implement it in a low level and efficient programming language.
In the current PoC $\proofGenerator$ uses \project{libsnark} and is implemented in \project{C++}.
Alternatively, one could implement $\proofGenerator$ in \project{Rust} using \project{Bellman}\footnote{\url{https://github.com/zkcrypto/bellman}}, for instance.

Making changes in one of the modules is completely transparent for the others. That is, changes in $\proofGenerator$ do not impact $\proofConsumer$ in any way, provided the API is not changed and the format of requests and responses is the same, and vice versa.
Components are loosely coupled what makes the entire architecture more flexible.

\subsubsection{Using \zeth{} with ERC20/ERC223 tokens}
The pseudocode of the $\mix$ function, cf.~\cref{fig:mixcode}, is compatible to work with ERC20 or ERC223 tokens.
In fact, $\sendValue$ is a pseudocode function that represents the action of sending value from a contract to a recipient. (Here, the recipient of the funds is the sender of the transaction).
In order to use the mixer with a given digital asset, one needs to use the $\sendValue$ function bound to that asset. Namely, if Ether is used, the solidity code for $\sendValue$ is \lstinline{msg.sender.transfer(amount)}. However, if $\Mixer$ is used to transfer ERC20 tokens, the solidity code for $\sendValue$ is \lstinline{erc20TokenInstance.transfer(msg.sender, amount)}.

In addition to properly instantiate $\sendValue$, a few other minor changes need to be introduced in $\Mixer$.
In fact, the contract needs to be given the address of the ERC20 token to use and the right allowances need to be set for $\Mixer$ to be able to modify user balances on the token contract on their behalf.

\subsubsection{Ethereum events to push data to consumers}
In general, retrieving data from a database may cause data leakage and privacy issues. The sole fact that some user is querying a particular database may reveal private information.
For example, in the case presented in this paper, a user who constantly queries the mixer contract could be identified as a receiver of a payment who wants to fetch their new notes. In particular, if an adversary observing the mixer notices that a payment made by some party $\pdv$ is followed by queries to $\Mixer$ by another party $\pdv'$, they may conclude that the payment was sent from $\pdv$ to $\pdv'$.

In order to prevent any data leakage associated with users' queries, we leverage Ethereum events to log important data and push \zeth{}-related pieces of information directly to each subscriber.
This enables $\Mixer$ users to be aware of the current state of the contract without requiring them to query the blockchain and obfuscates the link between a sender and their recipients.

\subsubsection{Multiple payments in a single transaction}
The ability to emit multiple events in a single smart contract call makes it possible to perform multiple payments within a single transaction.
In fact, Alice can pay both Bob and some other party Charlie by generating a \zethNote{} for each of them, and then encrypting each note with their encryption keys.
The call to $\mix$ broadcasts both ciphertexts using Ethereum events. The recipients can retrieve their \zethNotes{} by decrypting them.
The upper bound for the number of parties that can be paid in a single transaction is given by the parameter $\coinsOut$ in the joinsplit equation (cf.~\cref{eq:joinsplit}).

The possibility to do multiple payments in a single transaction leads to some hope regarding scalability of the proposed protocol as it could be used to limit the number of transactions submitted to the system.
Unfortunately, increasing the number of input notes $\coinsIn$ significantly affects the performance of the system as checking multiple Merkle path efficiently remains a major bottleneck.
Moreover, increasing $\coinsOut$ while keeping $\coinsIn$ small is not a viable solution. In fact, a large number of output notes $\coinsOut$ could lead to plethora of notes of negligible value. Making any payment with such scattered funds would require multiple transactions to gather them together.
Further work is required in this direction.

\subsubsection{Gas cost}
The overall gas cost of a $\mix$ call is dominated by the gas cost of a zkSNARK proof $\pi$ verification. (For the details of the underlying zkSNARK $\snark$ used in the PoC check \cite[Fig.~10, App.~B]{DBLP:journals/iacr/Ben-SassonCTV13}.) We express the verification cost as a function of the gas cost of key operations such as point addition ($\ECADDGas$), scalar multiplication ($\ECMULGas$) and pairings equality checks ($\PairBaseGas$ for each equality to check and additionally $\PairPerPointGas$ for each pairing to compute\footnote{Note that the gas cost of the group operation in the target group of the bilinear pairing is included in $\PairPerPointGas$ along with the computation of the bilinear pairing itself. Thus, it is not necessary to account for it separately in the gas cost computation.}.) on $\texttt{bn256}$.

\newcommand{\IC}{\mathsf{IC}}
\newcommand{\GRP}{\mathbb{G}}
\newcommand{\pair}[2]{e(#1, #2)}
\newcommand{\ggen}{\mathfrak{g}}
Let $\inp = (x_1, \ldots, x_n)$ denote the instance and $\smallset{\vk_{\IC, i}}_{i = 0}^{n}, \allowbreak \vk_{A}, \allowbreak \vk_{B}, \allowbreak \vk_{C}, \allowbreak \vk_{\gamma}, \vk^{2}_{\beta \gamma}$ represent the CRS elements all belonging to group $\GRP_1$ or $\GRP_2$;
proof $\pi$ consists of $8$ elements: $\pi_A, \pi'_A, \pi'_B, \pi_C, \pi'_C, \pi_K, \pi_H \in \GRP_1$ and $\pi_B \in \GRP_2$.
Similarly to~\cite{DBLP:journals/iacr/Ben-SassonCTV13} we define groups $\GRP_1, \GRP_2, \GRP_T$ of size exponential to the security parameter $\secpar$ and a bilinear pairing $e: \GRP_1 \times \GRP_2 \to \GRP_T$. We denote group's $\GRP_i$ generator by $\ggen_i$ and use additive notation for $\GRP_1$ and $\GRP_2$, and multiplicative notation for $\GRP_T$.
Operations performed by a verifier contract $\Verifier$ follows $\snark.\verifier$ and come with the costs:
\begin{enumerate}
  \item Compute the linear combination
  \[
    \vk_{\vec{x}} \gets \vk_{\IC, 0} + \sum_{i = 1}^{n} x_i \cdot \vk_{\IC, i} \in \GRP_1\enspace,
  \]
  gas cost: $n \cdot (\ECMULGas + \ECADDGas) + \ECADDGas$.
  \item Check the validity of knowledge commitments:
  \begin{align*}
    \pair{\pi_A}{\vk_A} & = \pair{\pi'_A}{\ggen_2}\enspace,\\
    \pair{\vk_B}{\pi_B} & = \pair{\pi'_B}{\ggen_2}\enspace,\\
    \pair{\pi_C}{\vk_C} & = \pair{\pi'_C}{\ggen_2}\enspace,
  \end{align*}
  gas cost: $3 \cdot (\PairBaseGas + 2 \cdot \PairPerPointGas)$.
  \item Check that the same coefficients were used:
  \[
    \pair{\pi_K}{\vk_\gamma} = \pair{\vk_{\vec{x}} + \pi_A + \pi_C}{\vk^{2}_{\beta \gamma}} \cdot \pair{\vk^{1}_{\beta \gamma}}{\pi_B}\enspace,
  \]
  gas cost: $\PairBaseGas + 3 \cdot \PairPerPointGas + 2 \cdot \ECADDGas$.
  \item Check QAP divisibility:
  \[
    \pair{\vk_{\vec{x}} + \pi_A}{\pi_B} = \pair{\pi_H}{\vk_Z} \cdot \pair{\pi_C}{\ggen_2}\enspace,
  \]
  gas cost: $\PairBaseGas + 3 \cdot \PairPerPointGas + \ECADDGas$.
\end{enumerate}

Using the Ethereum parameters\footnote{\url{https://github.com/ethereum/go-ethereum/blob/master/params/protocol_params.go}} and the fee schedule described in the Ethereum Yellow Paper~\cite{yellow-paper}, we can estimate the gas cost of an on-chain proof verification.
The overall gas cost associated with the verification of~\cite{DBLP:journals/iacr/Ben-SassonCTV13} zkSNARK lies below $2$ million gas when the joinsplit equation (cf.~\cref{eq:joinsplit}) has $2$ private inputs and $2$ private outputs. It is worth mentioning that this cost is expected to diminish tremendously if EIP1108\footnote{\url{https://github.com/ethereum/EIPs/blob/master/EIPS/eip-1108.md}} gets accepted and implemented.
Importantly, modifying the number of inputs and outputs supported by the joinsplit equation increases the size of the circuit which affects the proving time. In addition, a larger instance increases the verification time, the verification cost, and the number of modifications made in the Ethereum storage at each function call.
While verifying the zkSNARK proof constitutes the main gas cost of the $\mix$ function, it is important to remember that storing value on Ethereum is expensive as well. As a consequence, deploying a mixer contract maintaining a very deep Merkle tree will not come for free, as the entire set of leaves\footnote{number of leaves = $2^{\variable{tree\_depth}}$} needs to be kept in the storage of $\Mixer$, along with all the serial numbers.
Keeping this is mind is essential when modify the parameters of the system.

\subsubsection{Further improvements}

As explained above, the piece of software implemented to test \zeth{} on Ethereum is a proof of concept resulting from the adaptation of existing projects. We emphasize that tremendous work is needed to have an acceptable proving time and a secure implementation that could be considered as a production-ready piece of software.
The Zcash team has done astounding work the past couple years and their \project{Sapling} network upgrade has led to a dramatic reduction of the complexity of their circuit, leading to significant optimizations and reduction in execution time.
Similar work could be carried out to improve \zeth{}, but this is left as a future work.

Furthermore, the current RPC interface can easily be extended to meet users' needs. The use of \project{gRPC} and of \project{Protocol\,Buffers} makes it trivial to define new services and extend existing ones. Extending the RPC interface and facilitating user experience is out of the scope of this work and let as a future work to the interested reader.

Last but not least, further work around the use of a multi-party computation ceremony is required to lessen trust assumptions around the generation of the CRS.

\section*{Acknowledgements}
We thank Kobi Gurkan for helpful comments on the first version of the paper.

\bibliographystyle{alpha}
\bibliography{references}

\appendix
\section{Beyond \ETHone{}}\label{sec:beyond-eth-one}
In the current version of Ethereum, Externally Owned Accounts (EOA) are secured via the use of ECDSA signature scheme~\cite{ECDSA} and a nonce mechanism. However, several proposals have been made to abstract these security mechanisms and allow for alternative protection systems. The end goal is to enable users of the network to define their own security model.

To do so, owners of EOA could provide their own transactions validation process, defining their own signature scheme and its verification code. That is, a transaction initiated by user's address $A$ is signed by $A$'s signature algorithm and verified using $A$'s verification code.
While the signature algorithm and the corresponding verification code can be picked arbitrarily by the owner of the account, the protocol should only allows verification procedures that could be executed succinctly enough to protect the system liveness. To that end, an upper bound of $50\,000$ gas for the verification function has been initially proposed in~\cite{on-abstraction} which has been increased to $200\,000$ in further proposals~\cite{on-abstraction-2} or~\cite{on-abstraction-3}.

Abstracting out the verification process of each transaction opens new possibilities for mixers (cf.~\cite{on-abstraction-4} for the case of ring signature mixers).
If we could define a verification function to verify a zkSNARK proof efficiently enough, we could enhance the protocol to enable calls to the $\mix$ function from newly generated addresses, possibly \emph{stealth address} with no funds, while ensuring that the miner would receive fees directly from the funds held on the mixer.
For example in case of \zeth{}, if arbitrary verification codes were supported, a party $\pdv$ willing to execute $\Mixer.\mix$ could first create a stealth address $\pdv'$, and then submit a~transaction $\tx$ to $\Mixer$ from $\pdv'$.
Since nobody should be able to connect $\pdv$ and $\pdv'$, the anonymity of $\pdv$ would be protected\footnote{We note that a similar approach was proposed independently in \cite{zether}}.


Nevertheless, the design of account abstraction is still under discussion and several trade-offs need to be taken into consideration to allow for more abstraction while keeping the system sound and secure~\cite{on-abstraction-5}.

\section{Security guarantees}\label{sec:security-guarantees}
The security guarantees of our scheme are inherited directly from the building blocks we use, especially Zerocash. The Zerocash authors described three necessary conditions for a payment system DAP to be private and secure. That is,
\begin{enumerate}
  \item ledger indistinguishability,
  \item transaction non-malleability, and
  \item balance.
\end{enumerate}
We briefly remind these notions. However, since our ambition is not to build a new ledger, but a mixer that works on top of some existing blockchain we shall define these properties in regard to a smart contract. Thus, in the following we talk about
\begin{inparaenum}[(1)]
  \item mixer indistinguishability,
  \item mixer transaction non-malleability, and
  \item mixer balance.
\end{inparaenum}
We briefly introduce these notions and debate about their importance.

\subsection{Mixer indistinguishability}\label{sec:mixer-indistinguishability}
  This property ensures that a malicious user cannot acquire information, other than publicly known data, from the smart contract. Looking at the protocol presented in~\cref{sec:overview-protocol}, public data could be, e.g.,~the account depositing money in the mixer, the amount deposited, the account transferring funds via the mixer, the account withdrawing funds along with the amount withdrawn.
  As mentioned, all these pieces of information are publicly accessible because users need to pay for the gas associated with the execution of these state transitions. Nevertheless, using \zeth{} obfuscates the transaction graph, cf.~\cref{fig:transaction-graph}, and pieces of information such as the recipients or the amount of a given transaction are hidden.

  As in~\cite{zerocash-paper}, mixer indistinguishability is defined as a game between an adversary $\adv$ and a challenger $\cdv$. The challenger initiates two mixers $\Mixer_0$ and $\Mixer_1$ and tosses a coin $b \sample \bin$. Then it sets $\Mixer_{left} = \Mixer_b$ and $\Mixer_{right} = \Mixer_{1 - b}$. $\adv$ is given access to $(\Mixer_{left}, \Mixer_{right})$ and is allowed to execute queries on both mixers. After querying the mixers, $\adv$ has to guess whether $b = 0$ or $b = 1$.

  Denote by $\ADDR$ a set of address key pairs, and by $\NOTE$ a set of \zethNotes{}. With this in mind, we present below the set of queries that can be submitted by $\adv$ to the challenger $\cdv$\footnote{The queries structure comes from Zerocash with only minor modifications to make them compatible with our mixer.}
(we describe each query with the associated steps performed by the challenger):

  \begin{description}
    \item $Q = (\createAddressAdv)$
    \begin{itemize}[--]
      \item Compute $\addr_{\pk}, \addr_{\sk}$;
      \item $\ADDR \gets \ADDR \cup \smallset{(\addr_{\pk}, \addr_{\sk})}$ \commenti{Note that $\ADDR$ differs from $\AddressRegistry$ since we also include $\addr_\sk$ in addition to $\addr_\pk$};
      \item return $\addr_{\pk}$.
    \end{itemize}

    \item $Q = (
    	\mixAdv,
    	\smallvec{\cmAddr^{\old}_i}_{i = 1}^{\coinsIn},
    	\smallvec{v_i}_{i = 1}^{\coinsOut},
    	v_{\vin},
    	v_{\vout},
    	\smallvec{\addr_{\pk, i}^{\old}}_{i = 1}^{\coinsIn},
    	\smallvec{\addr_{\pk, i}^{\new}}_{i = 1}^{\coinsOut})$
    \begin{itemize}[--]
      \item Compute root $\rt$ over all note commitments in $\MerkleTree$;
      \item For $i \in \range{1}{\coinsIn}$:
      \begin{itemize}[$\bullet$]
        \item let $\cm^{\old}_i$ be the commitment stored in address $\cmAddr^{\old}_i$ in $\MerkleTree$,
        \item find a note $\znote_i$ in $\NOTE$ that opens commitment $\cm_i^{\old}$,
        \item find $\addr_{\pk, i}^{\old}$ in $\ADDR$ that is $\znote_i$'s recipient address,
        \item compute a path $\mkPath^{\old}_{i}$ from $\cm_i^{\old}$ to $\rt$
      \end{itemize}
  		\item For $i \in \range{1}{\coinsOut}$:
  		\begin{itemize}
  			\item fetch from $\ADDR$ encryption key $k_{\pk, i}$ for the party identified with address $\addr^{\new}_{\pk, i}$,
  			\item create new note $\znote^{\new}_i$ using $a^{\new}_{\pk, i}$ from $\addr^{\new}_{\pk, i}$ and $v_i$,
  			\item encrypt note $\znote^{\new}_i$ using $k_{\pk, i}$, i.e.~compute $\cxx_i = \enc_{k_{\pk, i}} (\znote^{\new}_i)$.
  		\end{itemize}
  	  \item Compute proof $\pi$.
      \item Execute $\mix(\mixinput)$, abort if the function aborts.
      \item Add notes $\smallset{\znote^{\new}_i}_{i = 1}^{\coinsOut}$ to $\NOTE$.
    \end{itemize}
    \item $Q = (\receiveAdv, \addr_{\pk})$ \commenti{Tells the party that should receive money to update their status.}
    \begin{itemize}[--]
      \item Run $\receive$ on behalf of party $\pdv$ identified by $\addr_{\pk}$.
      \item Let $\smallset{\znote_i}_{i = 1}^{\coinsIn}$ be a set of notes output by $\receive$ in the previous step, then $\NOTE \gets \NOTE \cup \smallset{\znote_i}_{i = 1}^{\coinsIn}$.
      \item For $\smallset{\cm_{i}}_{i = 1}^{\coinsIn}$ being commitments obtained by $\pdv$ corresponding to $\smallset{\znote_i}_{i = 1}^{\coinsIn}$, output $\smallset{\cm_{i}}_{i = 1}^{\coinsIn}$.
    \end{itemize}
    \item $Q = (\insertAdv, \txmix)$ \commenti{Since the adversary can compute a transaction in their head and simply execute it on $\Mixer$.}
    \begin{itemize}[--]
			\item Execute $\mix (\txmix)$ and abort if it aborts.
      \item Run $\receiveAdv$ for all $\addr_{\pk}$ in $\ADDR$.
    \end{itemize}
  \end{description}

  In order to play the game, the adversary submits the queries in pairs $(Q, Q')$. If $\adv$'s query is $\insertAdv$, then $Q$ is executed on $\Mixer_{left}$ and query $Q'$ is executed on $\Mixer_{right}$. In the case of other query types, $\cdv$ executes $Q$ on $\Mixer_0$ and executes $Q'$ on $\Mixer_1$. Both queries have to be \emph{publicly consistent}, i.e.~be of the same type and should share the public input\footnote{We refer the reader to~\cite{zerocash-paper} for a formal definition of publicly consistent queries.}. For $(a_0, a_1)$ being the output of the submitted queries, $\cdv$ sends it back to $\adv$ in the form of a tuple $(a_b, a_{1 - b})$. At the end of the game, $\adv$ needs to guess the value $b$ initially chosen by $\cdv$.

   It is worth noting that the set of queries described in the game above allows $\adv$ to modify the mixers' internal storage via the execution of $\mix$ that triggers payments from some parties $\setsender$ to some other parties $\setrecipient$ (even though $\adv$ does not know the senders' secret keys).

  Mixer indistinguishability of \zeth{} follows directly from the ledger indistinguishability of Zerocash.

\subsection{Transaction non-malleability}\label{sec:transaction-non-malleability}

Transaction non-malleability assures that no adversary $\adv$ can use a transaction $\txmix$ to produce another transaction $\txmix^{*}$ that uses notes leading to the same serial numbers and that is acceptable from a state that has been obtained by a sequence of transitions that does not contain $\txmix$ (note that $\txmix^{*}$ cannot be accepted by a smart contract which state has been obtained by the application of $\txmix$ since the serial numbers revealed in $\txmix$ and $\txmix^{*}$ are the same). To illustrate the importance of non-malleability we show a possible misuse of the system if this property does not hold.

Assume that an adversary $\adv$ has an overview of the whole Ethereum network and observes a transaction $\txmix$ (note that Ethereum transactions do not immediately reach all the network nodes. Propagating data, i.e.~transactions, blocks, to all the nodes takes some time). Let us denote by $P$ the part of the network on which $\adv$ has seen the transaction $\txmix$, and by $P'$ its complement. After observing $\txmix$ on $P$, $\adv$ can generate another transaction $\txmix^{*}$, related to $\txmix$, and disseminate it in $P'$. If $P'$ propagates faster, then $\txmix^{*}$ could be accepted and $\txmix$ could be rejected.
Denote by $\recipient$, resp.~$\recipient^{*}$, the recipient of funds in transaction $\txmix$, resp.~$\txmix^{*}$. If the recipients are different and $\txmix^{*}$ goes through, it means that $\adv$ managed to steal funds intended for $\recipient$ by passing them to $\recipient^{*}$.

Non-malleability is defined by a game $\TRNM$, cf.~\cref{fig:trnm}. More concretely, we say that a transaction scheme is non-malleable if for every NUPPT adversary $\adv$ and sufficiently large $\secpar$
\[
  \advantage{\TRNM}{\adv,\TRNM} \leq \negl \enspace.
\]
Informally, we say that $\adv$ wins the game if they manage to make a transfer transaction $\txmix^*$, such that there exists another mix transaction $\txmix$, that reveals the same serial number as $\txmix^{*}$, yet $\mix_{\stateMixer'}(\txmix^{*})$ does not abort. Here $\mix_{\stateMixer'}$ represents the $\mix$ function being executed on the mixer on state $\stateMixer'$, i.e.~the state of the mixer before transaction $\txmix$ was accepted.

\begin{figure}
  \centering
  \fbox{
  \procedure[linenumbering]{$\TRNM$}{
    \txmix^{*} \gets \adv^{\oledger}\\
    \pcreturn \exists \txmix \in \Mixer: (\txmix.\sn = \txmix^{*}.\sn) \land (\txmix \neq \txmix^{*}) \land (\mix_{\stateMixer'}(\txmix^{*}) \neq \bot)
  }
  }
  \caption{Transaction non-malleability game.
  Denote a serial number $\sn$ revealed by transaction $\txmix$ by $\txmix.\sn$.
  The adversary $\adv$ is given oracle access to ledger $\ledger$, where a smart contract $\Mixer$ is deployed. $\adv$ wins if they manage to produce a mixer transaction $\txmix^{*}$ such that there is another transaction $\txmix$ in $\Mixer$ that has the same serial number but function $\mix_{\stateMixer'}$ that also verifies transactions does not abort, for a mixer state $\stateMixer'$ just before $\txmix$ was included.}
  \label{fig:trnm}
\end{figure}

Intuitively, transaction non-malleability of our scheme comes directly from the non-malleability of the underlying zkSNARK and $\INDCCA$-security of the encryption scheme. Recall the transaction syntax $\txmix = (\mixinput)$, where the first part, i.e.~$\verifyinput$, is a zero-knowledge proof and the statement it proves (call it \emph{proof part}), and the last tuple $\smallvec{\cxx_i}_{i = 1}^{\coinsOut}$ is an encryption of the transferred \zethNotes{} (call it \emph{ciphertext part}).
Now, if $\adv$ were able to produce $\txmix^{*}$ different from $\txmix$ but with the same serial numbers, then $\txmix^{*}$ and $\txmix$ would either differ on their proof part or on their ciphertext part.
First, note that $\adv$ cannot change any of the ciphertexts $\smallvec{\cxx_i}_{i = 1}^{\coinsOut}$ without changing the underlying plaintexts, since $\INDCCA$ holds.
Likewise, any change in the plaintext changes the statement for the zero-knowledge proof and requires to provide a new proof $\pi'$. Since the zkSNARK used is simulation-extractable, any adversary who provides proof $\pi$ for an instance $\inp$, such that a pair $(\inp, \pi)$ has not been computed before, has to know a corresponding witness $\wit$.

\subsection{Balance}\label{sec:balance}
The balance property ensures that no adversary $\adv$ can make money ``out of thin air". To capture this intuition,~\cite{zerocash-paper} defines a game $\balance$ in which $\adv$ wins if they own or spend more funds that they have either deposited or received from other parties via $\Mixer$. Following the notation in the above-mentioned games, we say that a smart contract is balanced if for every PPT adversary $\adv$ and sufficiently large $\secpar$
\[
  \advantage{\balance}{\adv,\balance} \leq \negl\enspace.
\]

\renewcommand{\setznotes}{\smallset{\znote_i}_{i \in I}}

At first a mixer $\Mixer$ is set up by a challenger $\cdv$ and $\adv$ is given oracle access to it. $\adv$ is free to issue queries to $\Mixer$ as in \cref{sec:mixer-indistinguishability}.
At the end of the game, $\adv$ sends $\cdv$ a set of notes $\setznotes$, for some set $I$. Challenger $\cdv$ computes the following quantities:
\begin{itemize}[$\bullet$]
  \item $\val^{\variable{unspent}}$: the total value of spendable notes in $\setznotes$. We call a note $\znote$ spendable if it is feasible to create a transaction $\txmix$ such that $\txmix$ uses $\znote$'s data as input and is acceptable by $\Mixer$.
  \item $\val^{\variable{publicIn}}$: the total value of public input $\val^{\vin}$ in transactions submitted by $\adv$.
  \item $\val^{\variable{publicOut}}$: the total value of public output $\val^{\vout}$ in transactions submitted by $\adv$.
  \item $\val^{\variable{inc}}$: value of all notes transferred to $\adv$ from addresses in $\ADDR$.
  \item $\val^{\variable{exp}}$: value of all notes transferred from $\adv$ to addresses in $\ADDR$.
\end{itemize}
Note that all values can be computed by $\cdv$ by observing $\adv$'s queries to $\Mixer$.
We say that $\adv$ wins the game $\balance$ iff
\[
  \val^{\variable{unspent}} + \val^{\variable{publicOut}} + \val^{\variable{exp}} > \val^{\variable{publicIn}} + \val^{\variable{inc}}\enspace.
\]
The proof of balance property follows the proof given in Zerocash. However, we note that $\prf^{\addr}$ needs to be collision resistant to fix the omission in Zerocash security proof, as mentioned in~\cite[Section 8.8]{zcash-protocol-spec}.
\end{document}